\documentclass[12pt]{scrartcl}

\usepackage[utf8]{inputenc}
\usepackage[T1]{fontenc}
\usepackage{cite}
\usepackage{pslatex}
\usepackage{graphicx}
\usepackage{color}
\usepackage{scalefnt} 
\def\LO{\mbox{\scriptsize LO}}
\def\NNLO{\mbox{\scriptsize NNLO}}
\def\QCD{\mbox{\scriptsize QCD}}
\def\EW{\mbox{\scriptsize EW}}

\def\TeV{\mbox{\scriptsize TeV}}

\newcommand{\be}{\begin{equation}}
\newcommand{\ee}{\end{equation}}
\newcommand{\bea}{\begin{eqnarray}}
\newcommand{\eea}{\end{eqnarray}}

\newcommand{\GeV}{{\rm GeV}}

\def\pt{p_{\rm T}}

\graphicspath{{./figures/}}

\begin{document}
\title{
\begin{flushright}
{\normalsize   TTP/13-015\\
  HU-EP-13/26\\
 October  2014\\}
\end{flushright}
\vspace{1.0cm}
\large Weak Interactions in Top-Quark Pair Production\\ 
at Hadron Colliders: An Update} 
\author{\small
   J.H.~K\"uhn$^a$,\,
   A.~Scharf\,$^{b}$\, and  
   P.~Uwer\,$^c$\\[3mm]
  $^a${\small {\em Institut f{\"u}r Theoretische Teilchenphysik,
  Karlsruhe Instititut of Technology (KIT),}}\\
  {\small {\em 76128 Karlsruhe, Germany}}\\[2mm]
  $^b${\small {\em 
Institut f\"ur Theoretische Physik und Astrophysik, 
Universit\"at W\"urzburg,}}\\ 
{\small {\em D-97074 W\"urzburg, Germany}}\\[2mm]
  $^c${\small {\em Institut f\"ur Physik,
  Humboldt-Universit\"at zu Berlin,}}\\
  {\small {\em 12489 Berlin, Germany}}
}
\date{}
\maketitle
\begin{abstract}
Weak corrections for top-quark pair production at hadron
colliders are revisited. Predictions for collider energies of 8~TeV,
adopted to the recent LHC run, and for 13 as well as 14~TeV, presumably relevant
for the next round of LHC experiments, are presented. Kinematic regions
with large momentum transfer are identified, where the corrections 
become large and may lead to strong distortions of differential 
distributions, thus mimicking anomalous top quark couplings. 
As a complementary case we investigate the threshold region, 
corresponding to configurations with small relative
velocity between top and antitop quark, which is particularly 
sensitive to the top-quark Yukawa coupling. We demonstrate, that
nontrivial upper limits on this coupling, complementary to those
recently derived by the CMS and the ATLAS collaorations, are well 
within reach of ongoing experiments. We, furthermore, suggest a
prescription that allows the implementation of these corrections in
current Monte Carlo generators. Furthermore, the weak corrections are
have been included in the publicly available Hathor library. The
numerical results presented in this article use the same setup as the
recently calculated NNLO QCD corrections. The results can thus be
combined to give the most precise theoretical predictions.
\\[2mm]
\end{abstract}


\section{Introduction}
\label{intr}
During the past years the determination of the top quark mass, 
its couplings, production and decay 
rates has been pursued successfully at the Tevatron. Based on an integrated
luminosity of almost  $10~{\rm fb}^{-1}$ per experiment collected by 
both CDF and D0 at 1.96~TeV, a sample of nearly 100000 top quark pairs has been 
produced. The analysis of these events 
has led, for example, to a top mass determination  
 of $M_t= 173.18 \pm 0.94 {\rm GeV}$ \cite{Aaltonen:2012ra}, 
corresponding to a relative error of about half percent. 
The total production cross section $\sigma_{t\bar t} = 7.65 \pm
0.42$~pb \cite{D0confnote6363} determined at Tevatron 
is in very good  agreement with the theory predictions
\cite{Moch:2008qy,Aliev:2010zk,Moch:2012mk,Cacciari:2011hy,Czakon:2012pz,Czakon:2012zr,Baernreuther:2012ws,Ahrens:2011px,Kidonakis:2011ca,Kidonakis:2008mu,Beneke:2012wb,Czakon:2013goa}.
The same is true for the cross section measurements performed at the
LHC \cite{Aad:2012vip,Aad:2012mza,Chatrchyan:2012ria,Chatrchyan:2013faa,Chatrchyan:2013ual,Chatrchyan:2013kff,Khachatryan:2014loa}.
Also the $t\bar t$ invariant mass distribution has
been measured at LHC over a wide kinematical range
\cite{:2012qa,:2012rq,Chatrchyan:2012saa,Chatrchyan:2012yca,Chatrchyan:2012cx,Aad:2013nca}. 
Similar to the cross section measurements the
results are in agreement with the Standard Model (SM) predictions.  
In contrast,
surprising deviations from the theory predictions  
have been observed in the Tevatron experiments \cite{Abazov:2012oxa,Abazov:2011rq,Aaltonen:2012it,Aaltonen:2011kc} by investigating the
so-called charge asymmetry predicted originally fifteen years ago
\cite{Kuhn:1998jr,Kuhn:1998kw}. (For discussions of theoretical
predictions in the context of the SM see, for example, 
Refs.~\cite{Kuhn:2011ri,Hollik:2011ps,Bernreuther:2012sx,Ahrens:2011uf,Almeida:2008ug}).

Although these are impressive achievements already now, expectations for
top quark physics at the LHC fly even higher. Based on integrated
luminosities close to $5~{\rm fb}^{-1}$ per experiment at 7~TeV,
the top mass has been determined in a combined analysis 
to $M_t= 173.3 \pm 1.4~{\rm GeV}$ \cite{ATLAS:2012coa} already now. 
(Tevatron and LHC results combined have even led to a determination with
an error below five permille, 
$M_t= 173.34 \pm 0.76~{\rm GeV}$ \cite{ATLAS:2014wva}.)
With an integrated luminosity of more than $20~{\rm fb}^{-1}$ per experiment
collected recently at 8~TeV, several million top-quark pairs per
experiment have been produced. 
The high energy run at 14~TeV with its expected integrated luminosity of 
$100~{\rm fb}^{-1}$ will deliver about $10^8$ top quark
pairs per experiment during the coming years. The LHC is,
obviously, a factory of top quarks, allowing for a precise determination
of their properties and their production dynamics in a large kinematic region. 
The large center of mass energy available at the LHC will thus be used to
investigate top production with partonic subenergies of several TeV and
thus explore the point-like nature of the heaviest of the fundamental
particles. On the theoretical side precise predictions valid at the
highest accessible energies are required. With the recently completed 
next-to-next-to leading order QCD predictions 
\cite{Czakon:2012pz,Czakon:2012zr,Baernreuther:2012ws,Czakon:2013goa} a major step has
been taken. However when it comes to ultimate precision or highest
energies weak corrections significantly affect predictions within the
Standard Model. Two kinematic regions are of particular interest: 
\begin{enumerate}
\item[{\it i.})] Hard scattering events with partonic subenergies $\hat s$ and
  momentum transfers  $|\hat t|$  and $|\hat u|$ ($\hat s, \hat
  u$ and $\hat t$ denote the partonic Mandelstam variables) far larger than $M_t$ are
  affected by large negative corrections. These may reach
  nearly twenty percent, affecting transverse-momentum and angular
  distributions, and might well mimic anomalous top quark
  couplings. These negative corrections --- if not taken into account in
  the theoretical predictions --- could also hide a possible rise of the
  cross section due to a heavy resonance.
\item[{\it ii.})] The rate for events very close to the production threshold, 
  with relative top-antitop velocity $\beta\leq M_H/M_t$ is enhanced
  by the exchange of the relatively light Higgs boson. This effect 
  can be approximately described by a Yukawa potential and is 
  reminiscent of Sommerfeld rescattering corrections.
\end{enumerate}

Weak corrections to top quark pair production have first been
been studied twenty years ago \cite{Beenakker:1993yr}.  The complete
results, where some deficiencies were corrected and the result
given in closed analytical form, can be found in
Refs.~\cite{Kuhn:2005it,Bernreuther:2005is} and
Refs.~\cite{Bernreuther:2006vg,Kuhn:2006vh} for quark- and
gluon-induced processes, respectively. Numerical results (which,
however, differed from those presented in
Refs.~\cite{Bernreuther:2006vg,Kuhn:2006vh} and were corrected later) have
been published in Ref.~\cite{Moretti:2006nf}. Purely electromagnetic
corrections, which can be handled separately from the weak
corrections, are evaluated in Ref.~\cite{Hollik:2007sw}.  
As a consequence of cancellations between the positive contributions
from $\gamma$g-fusion and negative corrections to $q\bar q$-annihilation
the combined effect amounts at most to -4\%, if one considers
$\pt$-values as high as 1.5~TeV. The impact on the $\sqrt{\hat s}$
distribution remains below 1\%. The details of these corrections are
strongly cut-dependent and we refer to Ref.~\cite{Hollik:2007sw} for details. 

In the present paper we refrain from repeating the somewhat lengthy 
analytical formulae for the weak corrections 
and concentrate on the physics implications. We also
update results previously obtained using modern parton distribution functions
(PDF's) and the most recent values for the input parameters.
In Section 3 we present a prescription which allows the combination of
electroweak and NLO QCD corrections in the framework of current Monte
Carlo generators. Subsequently, in Section 4, we study the impact of
enhanced Yukawa couplings on the threshold behaviour in more detail and
contrast these potential measurements with recent experimental limits on
the Higgs boson decay rate.

\section{Large momentum transfers}
\label{sec-large-momenta}
Before entering the detailed numerical discussion, let us recall the
basic qualitative aspects of weak corrections for the present
case.
\begin{figure}
  \begin{center}
    \includegraphics[height=2.5cm]{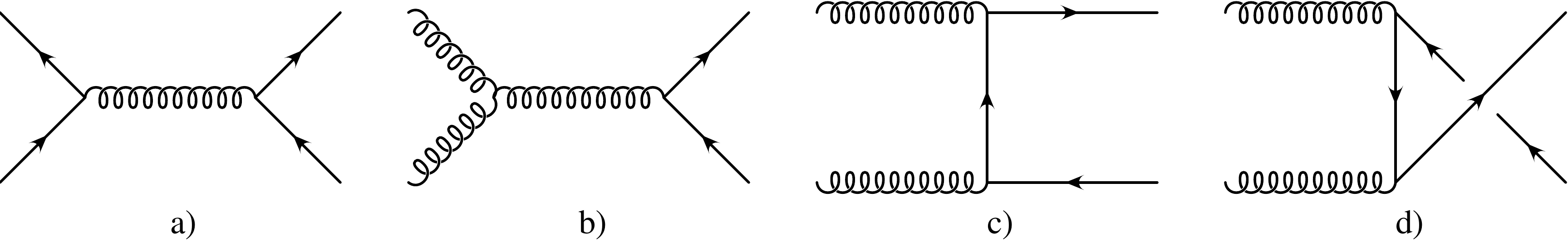}\vspace*{0.5cm}

    \includegraphics[height=2.5cm]{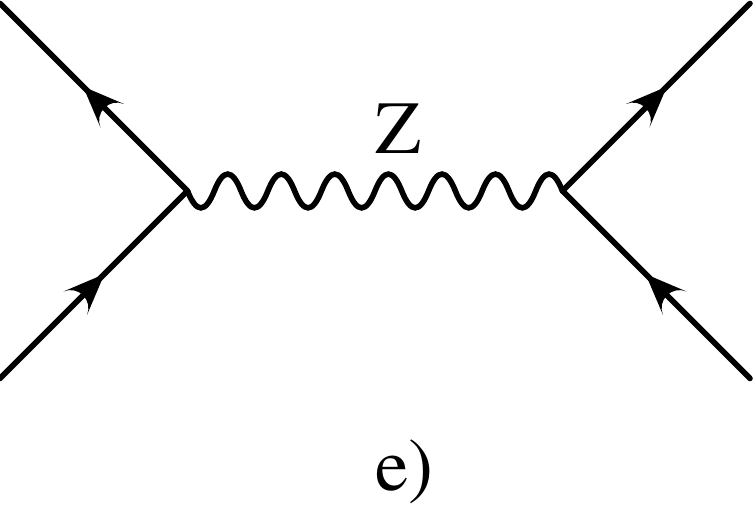}
    \caption{Lowest order QCD (a--d) and weak (e) amplitudes}
    \label{fig:Born}
  \end{center}
\end{figure}
With the Born amplitudes being of order
$\alpha_s$ (Figs.~\ref{fig:Born} a)-d) both for quark and gluon induced
QCD processes, and  of order $\alpha_{\rm weak}$ for
the lowest order weak process 
(Fig.~\ref{fig:Born} e), weak corrections start entering the cross
section at loop-induced order 
$\alpha_s^2\alpha_{\rm weak}$ only. The absence of an interference term
between the lowest order strong and neutral current amplitudes in the
quark induced process, which would be of order $\alpha_s\alpha_{\rm weak}$,
follows trivially from the different colour flow in the 
two relevant amplitudes Fig.~\ref{fig:Born}a and e, respectively.

\begin{figure}[!htbp]
  \begin{center}
    \leavevmode
    \includegraphics[height=6cm]{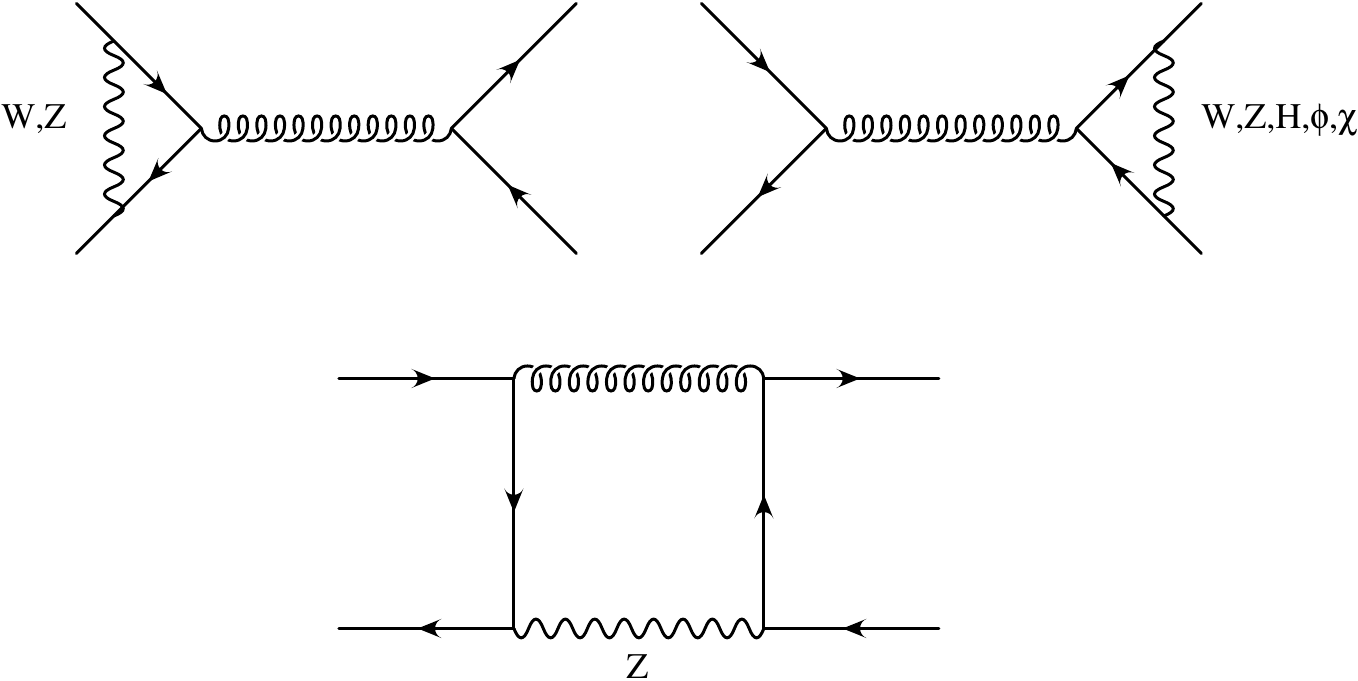}
    \caption{\label{fig:VirtualCor-qq}
      Sample diagrams for the virtual corrections to the quark-induced process.}
  \end{center}
\end{figure}

\begin{figure}[!htbp]
  \begin{center}
    \leavevmode
    \includegraphics[height=10cm]{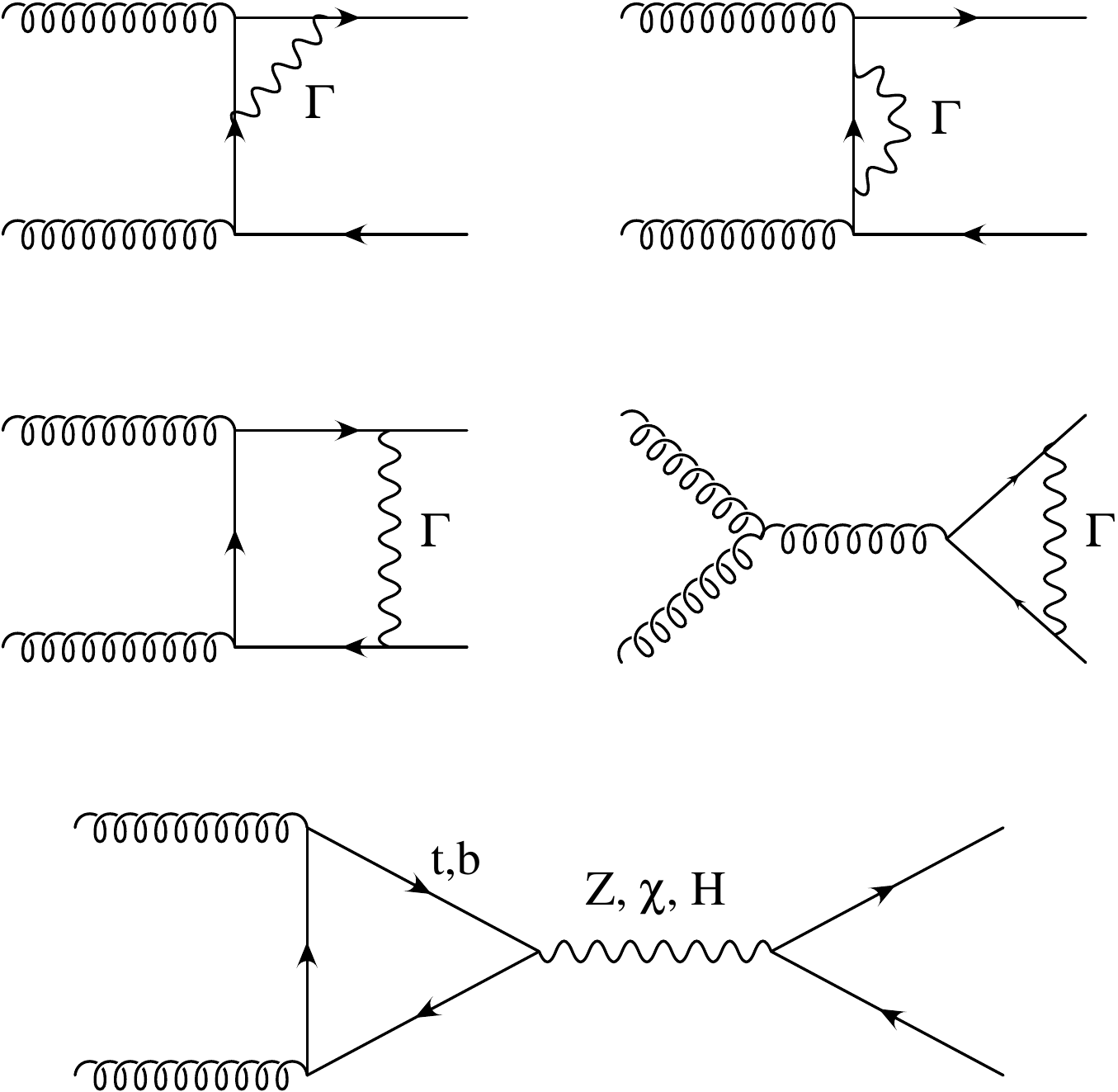}
    \caption{
          \label{fig:VirtualCor-gg}
          Sample diagrams for the virtual corrections for the
          gluon-induced process.  $\Gamma$
          stands for all contributions from gauge boson, Goldstone boson
          and Higgs exchange.}
  \end{center}
\end{figure}

Sample diagrams for weak corrections to quark- and gluon-induced
amplitudes using the 't Hooft-Feynman gauge are shown in Figs.~\ref{fig:VirtualCor-qq} and 
\ref{fig:VirtualCor-gg} ($\phi$ and $\chi$ denote the Goldstone bosons).
For gluon fusion weak effects start as
corrections to the QCD induced amplitudes. 

\label{sec:real}
\begin{figure}
  \begin{center}
    \includegraphics[height=6cm]{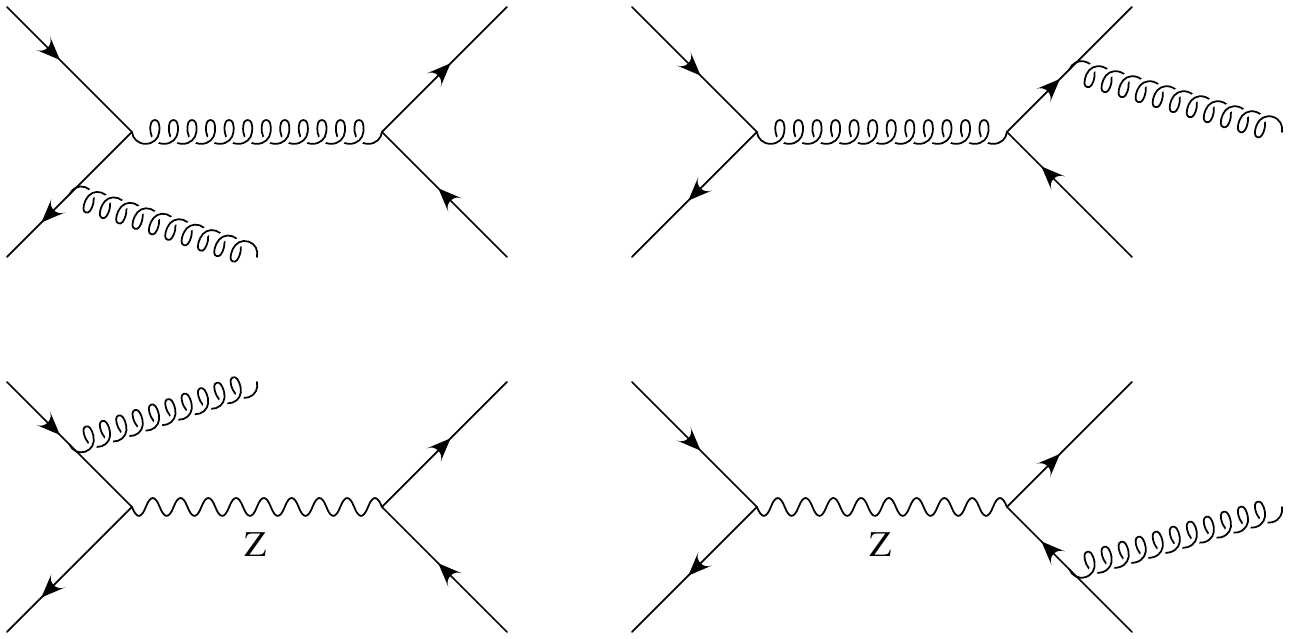}
    \caption{Sample diagrams for the real corrections to the
      quark-induced process.}
    \label{fig:RealCor}
  \end{center}
\end{figure}
For quark-antiquark annihilation the situation is more involved in view of
a specific class of order $\alpha_s^2\alpha_{\rm weak}$ 
contributions to the quark induced processes,  
which must be considered separately. In this case weak
and strong interaction are intimately intertwined, and corrections 
with virtual and real (Fig.~\ref{fig:RealCor}) 
gluon emission must be combined to arrive at an infrared finite result. 
The proper combination of real and virtual contributions is illustrated
in Fig.~\ref{fig:IR-cancellation}.
\begin{figure}
   \begin{center}
     \begin{picture}(100,80)
       \put(-140,-40){\includegraphics[width=9.cm]{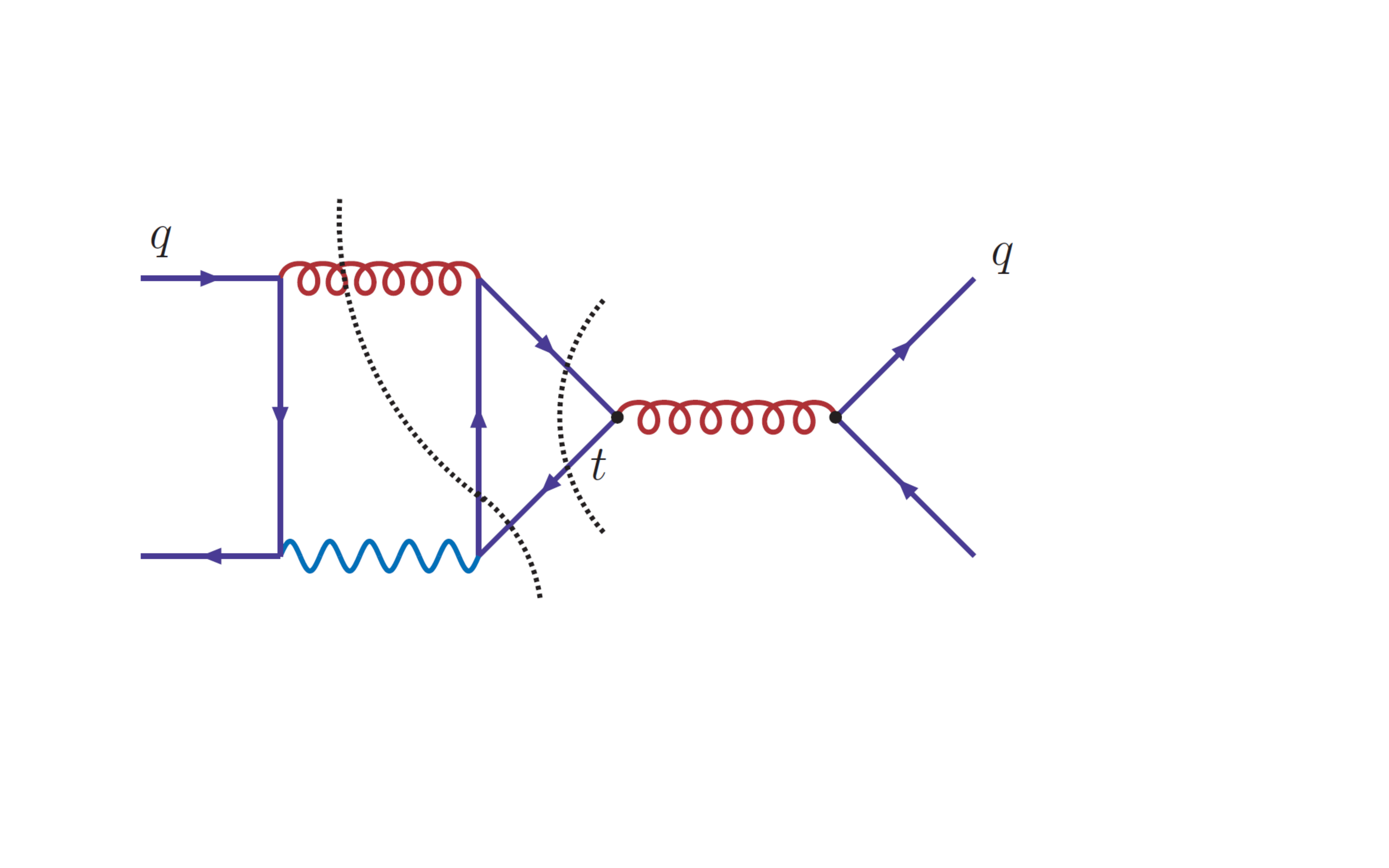}}
       \put(0,-40){\includegraphics[width=9.cm]{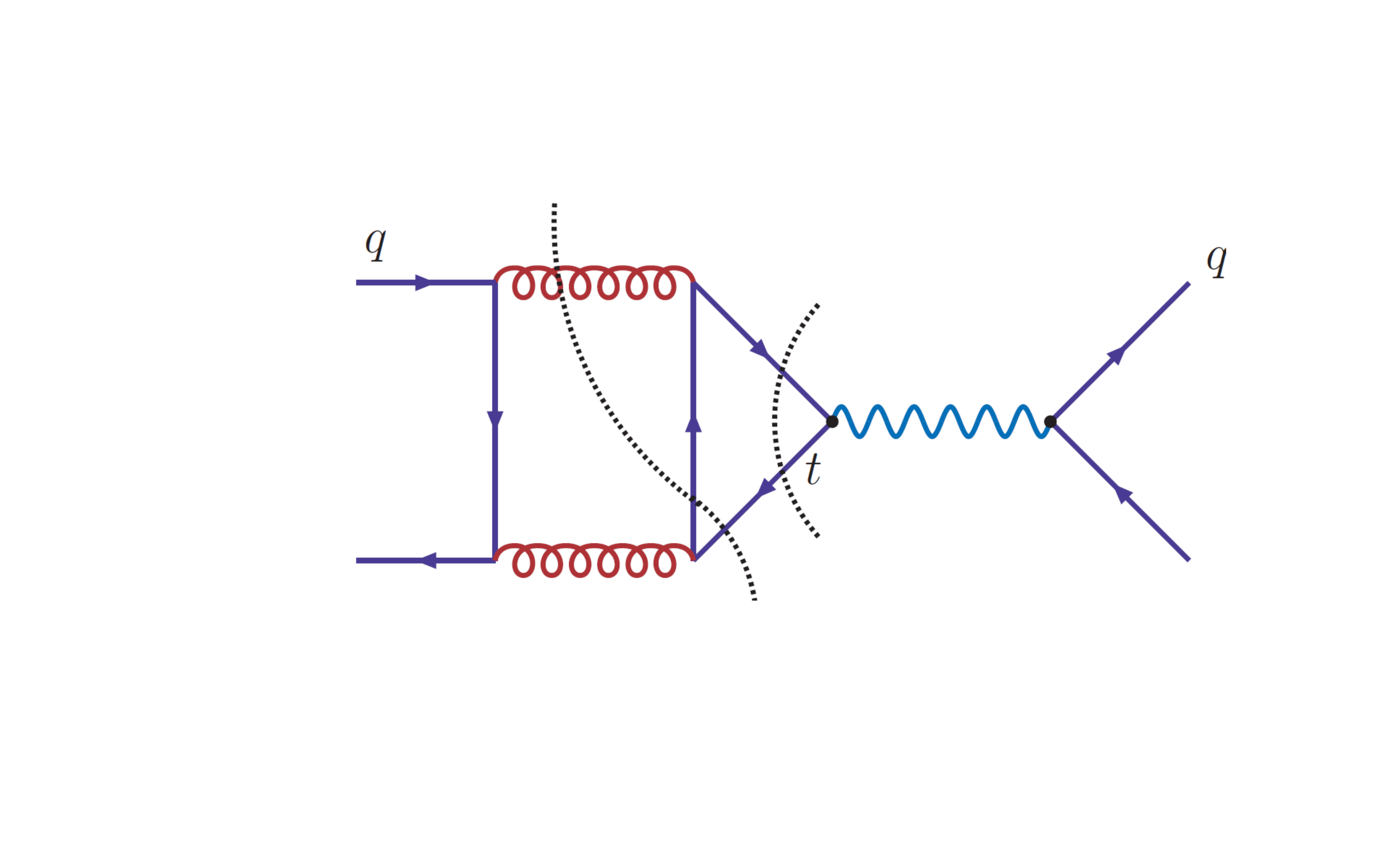}}
     \end{picture}
     \caption{Sample diagrams for the proper combination of virtual and real
       corrections to the quark-induced process. The dotted lines show
       different cuts corresponding to virtual and real corrections.}
     \label{fig:IR-cancellation}
   \end{center}
\end{figure}
This issue is discussed in more detail in Ref.~\cite{Kuhn:2005it}.
Only a specific combination of couplings is present in this case: The
top quark triangle in Fig.~\ref{fig:IR-cancellation} is attached to
two gluons with vector coupling. 
As long as we are interested in parity-even observables (like incl. cross
sections or $\pt$-distributions), the light quark coupling to the $Z$
boson is restricted to its axial coupling $g_A^q$
proportional to its isospin $I_3^q$.  This, in turn, leads to a strong
cancellation of this specific type of correction between u- and
d-quark induced processes. Since, furthermore, these contributions are
small (see Fig IV.3 of Ref.~\cite{Kuhn:2005it}) for one species of
quarks already, (less than one percent at threshold and about two
percent at very high energies), this group of corrections will be
neglected in the following discussion.  This observation might,
eventually, facilitate the combination of strong and weak corrections
discussed at the end of this paper.

For large parton energies the total corrections are negative, for quark- as
well as for gluon-induced processes. However, as a consequence of the
non-vanishing weak charge both in the initial as well as in the final
state, the corrections for quark induced top production are about twice 
those of the gluon induced process, with important consequences for the 
energy dependence of the corrections.

\begin{figure}
  \begin{center}
    \includegraphics[width=0.6\textwidth]{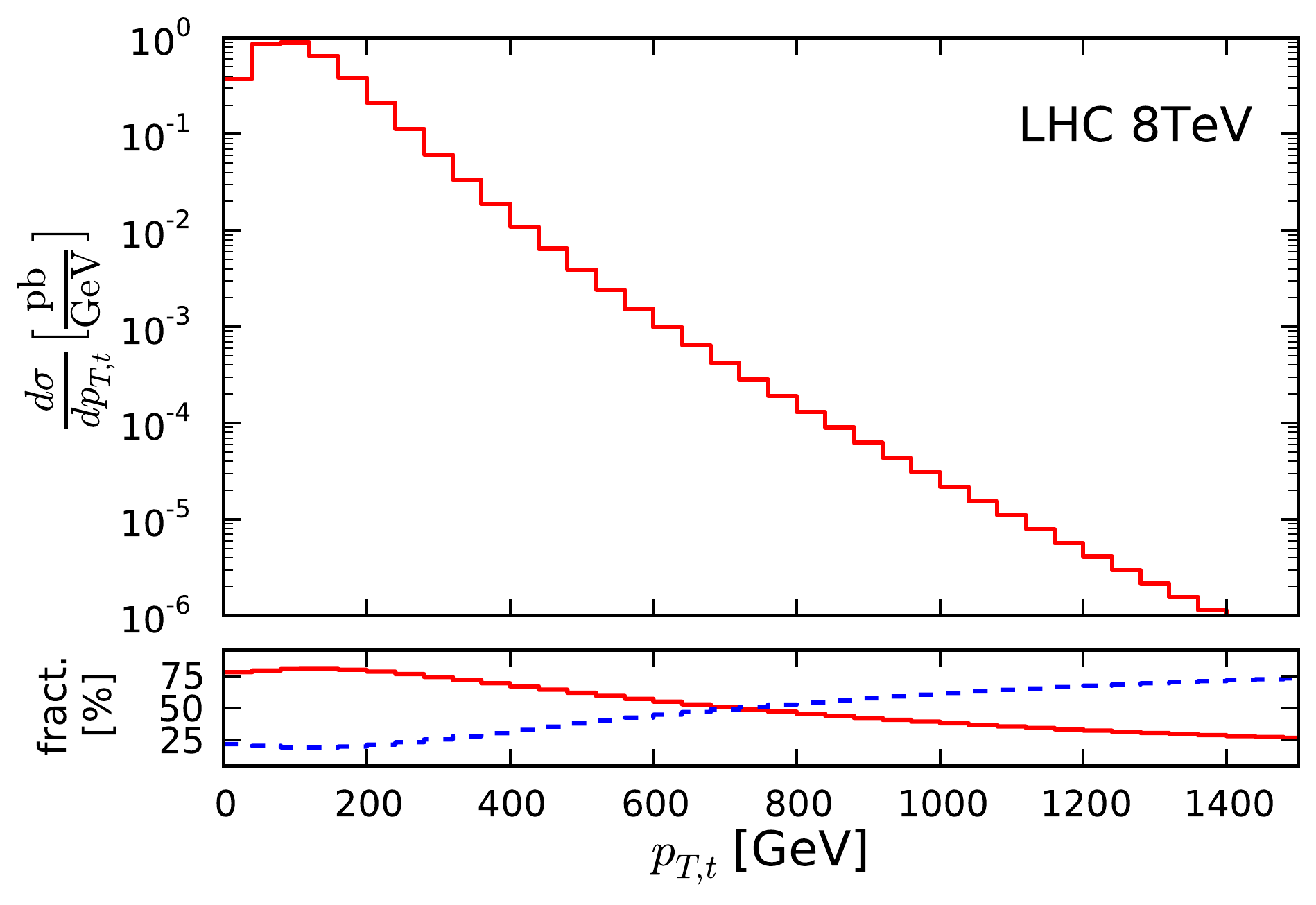}
    \caption{\label{fig:LHC8-pt}%
      Leading-order differential cross section for the LHC (8 TeV) 
      as a function of $\pt$. The lower plot shows the fraction from 
      gluon fusion (red,solid) and the fraction from 
      quark--antiquark annihilation (blue, dashed).
      }
  \end{center}
\end{figure}

As discussed in Ref.~\cite{Kuhn:2006vh} for proton-proton collisions at 14~TeV, 
the total cross section for top production
is dominated by gluon fusion. In contrast, production of top quarks with
at large transverse momenta 
is mainly induced by quark-antiquark annihilation, a consequence of the
different parton luminosities (see Fig.~\ref{fig:LHC8-pt} and \ref{fig:LHC8-mtt}
for LHC running at 8 TeV, results for LHC operating at 13 or 14 TeV
are given in appendix \ref{sec:13and14TeVresults}).
The relative increase of the the quark-induced processes in combination
with the different strength of
the weak corrections for the two reactions thus leads to an
additional increase of weak corrections for very large transverse 
momenta.
\begin{figure}
  \begin{center}
    \includegraphics[width=0.6\textwidth]{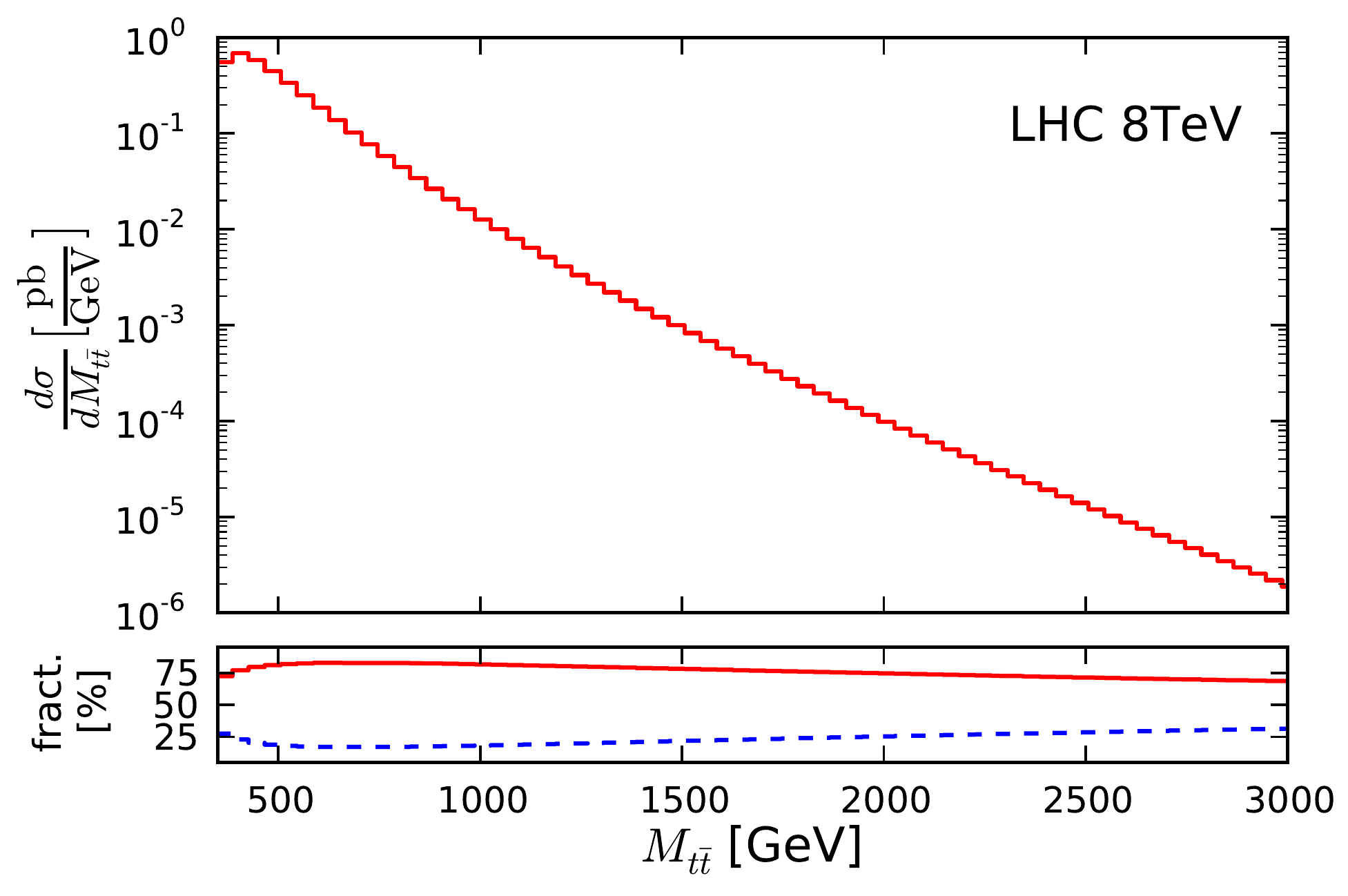}
    \caption{\label{fig:LHC8-mtt}%
      Leading-order differential cross section for the LHC (8 TeV) 
      as a function of $M_{t\bar t}$. 
      The lower plot shows the fraction from 
      gluon fusion (red,solid) and the fraction from 
      quark--antiquark annihilation (blue, dashed).
      }
  \end{center}
\end{figure}

For the numerical results presented
in this paper we use the parton distribution function 
MSTW2008NNLO PDF set\footnote{We follow closely the setup used for
  the NNLO QCD corrections \cite{Czakon:2012zr,Czakon:2012pz,Czakon:2013goa} 
so that the results presented
  here can be directly combined.} \cite{Martin:2009iq},
evaluated at a factorization scale $\mu_F=M_t$, and the coupling constants
\begin{displaymath}
  \alpha(M_t) = {1\over 127.0},\quad
  \alpha_s(M_t) = 0.106823,\quad
  \sin^2\theta_W  = 1-{M_W^2\over M_Z^2}.
\end{displaymath}
For the masses we use
\begin{displaymath}
  M_Z = 91.1876~\GeV,\quad
  M_W = 80.385~\GeV,\quad
  M_b = 4.82~\GeV,\quad
  M_t = 173.2~\GeV,
\end{displaymath}
and, if not stated otherwise, $M_H = 126~\GeV$.

Another important aspect is the nontrivial angular dependence of the
weak corrections. As is well known, the leading Sudakov
logarithms proportional $\log^2 (s/M_W^2)$ are only dependent on the
(weak) charge of the incoming and outgoing particles, 
subleading terms may exhibit a nontrivial angular dependence (see
e.g.\ \cite{Kuhn:1999nn,Kuhn:2001hz}). 
This is reflected in characteristic angular dependent 
virtual corrections which affect the rapidity distributions of top
quarks at the LHC and might well mimic anomalous couplings of the
particles involved.
\begin{figure}
  \begin{center}
    \includegraphics[width=0.45\textwidth]{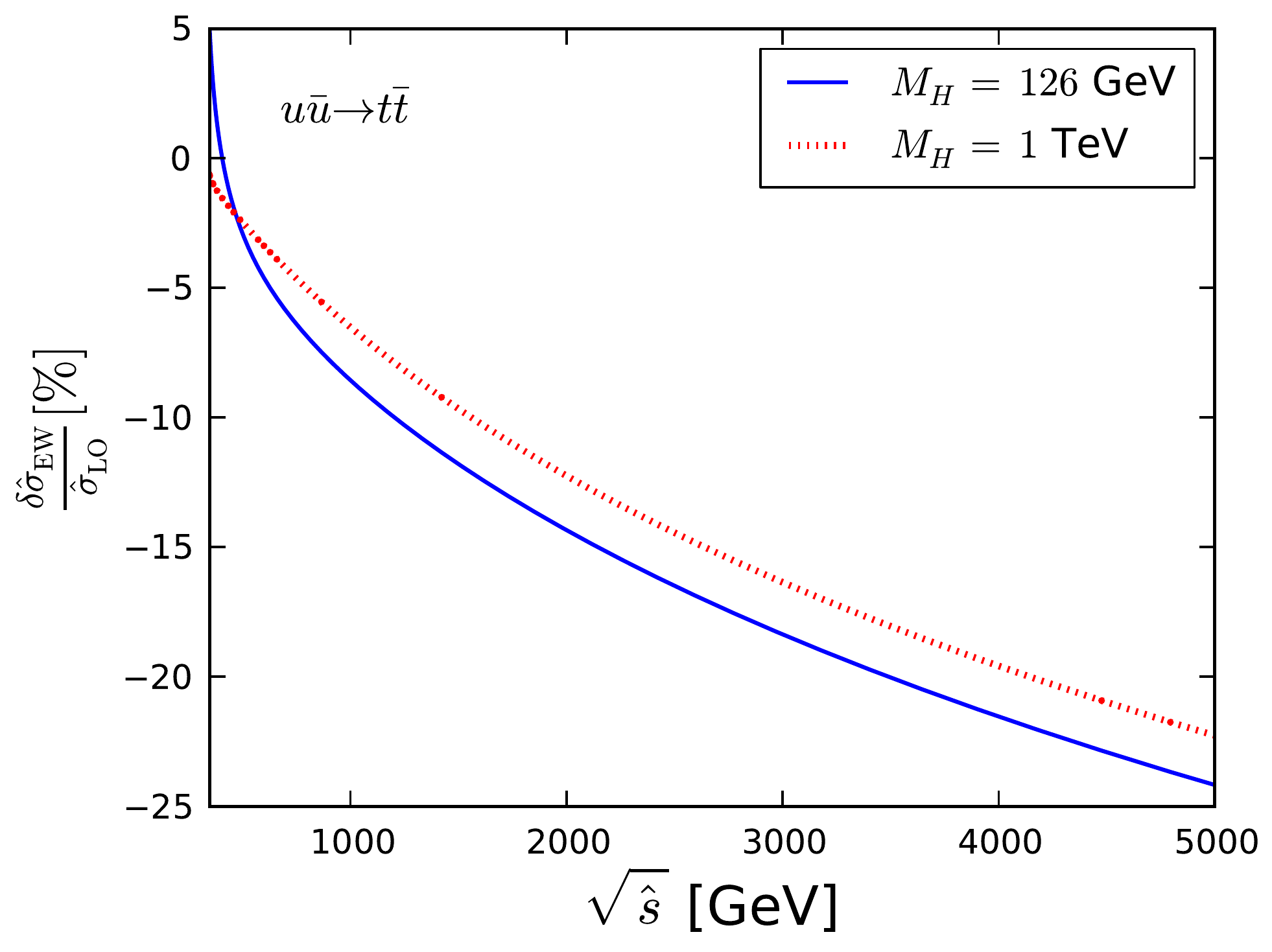}
    \includegraphics[width=0.45\textwidth]{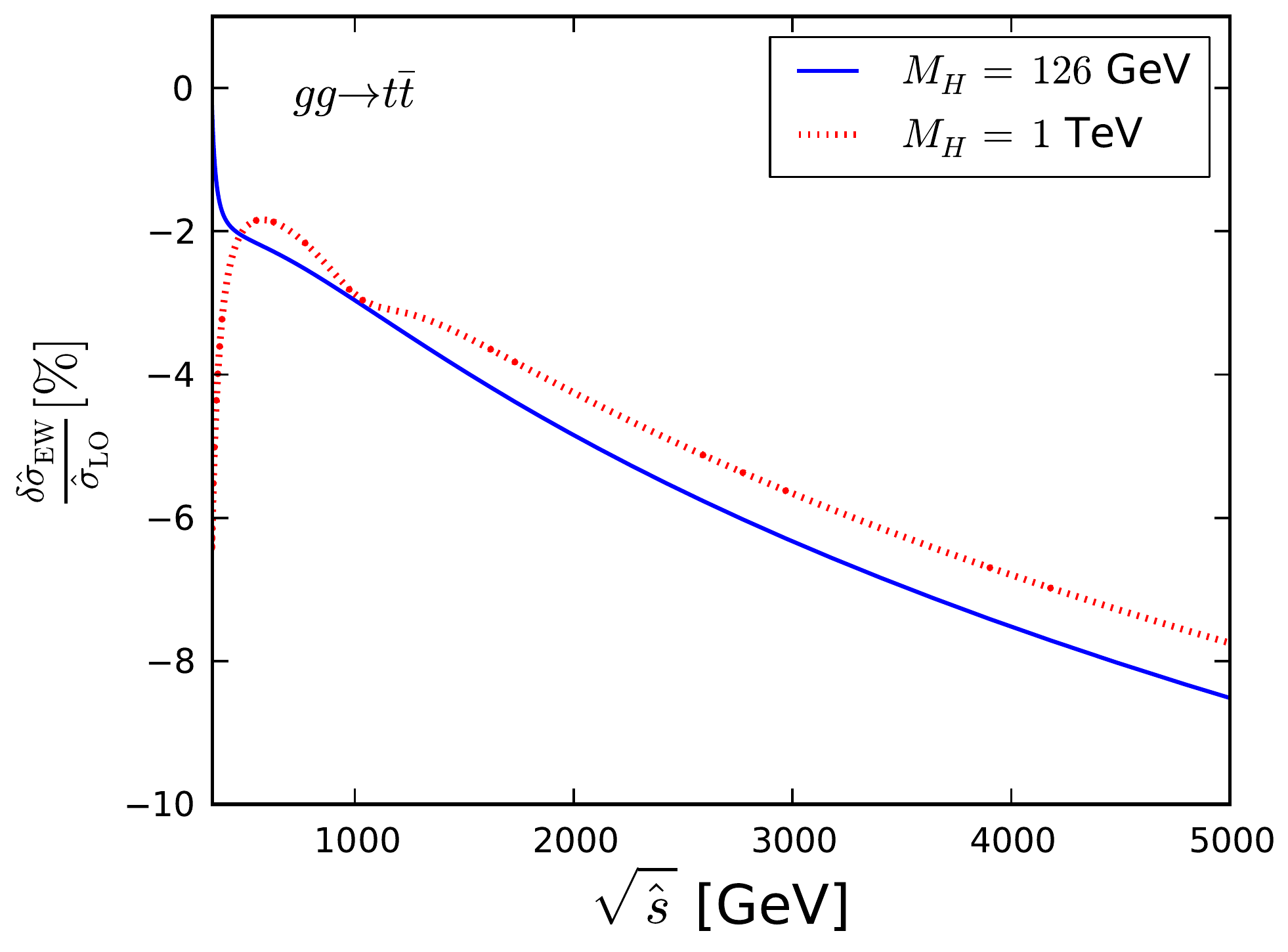}
    \caption{\label{fig:partoncor}Relative weak corrections at
  parton level for the quark- and gluon-induced reactions as functions
  of the squared parton energy $\hat s$ for two characteristic masses
  of the Higgs boson.}
     \end{center}
\end{figure}   
\begin{figure}
  \begin{center}
    \includegraphics[width=0.45\textwidth]{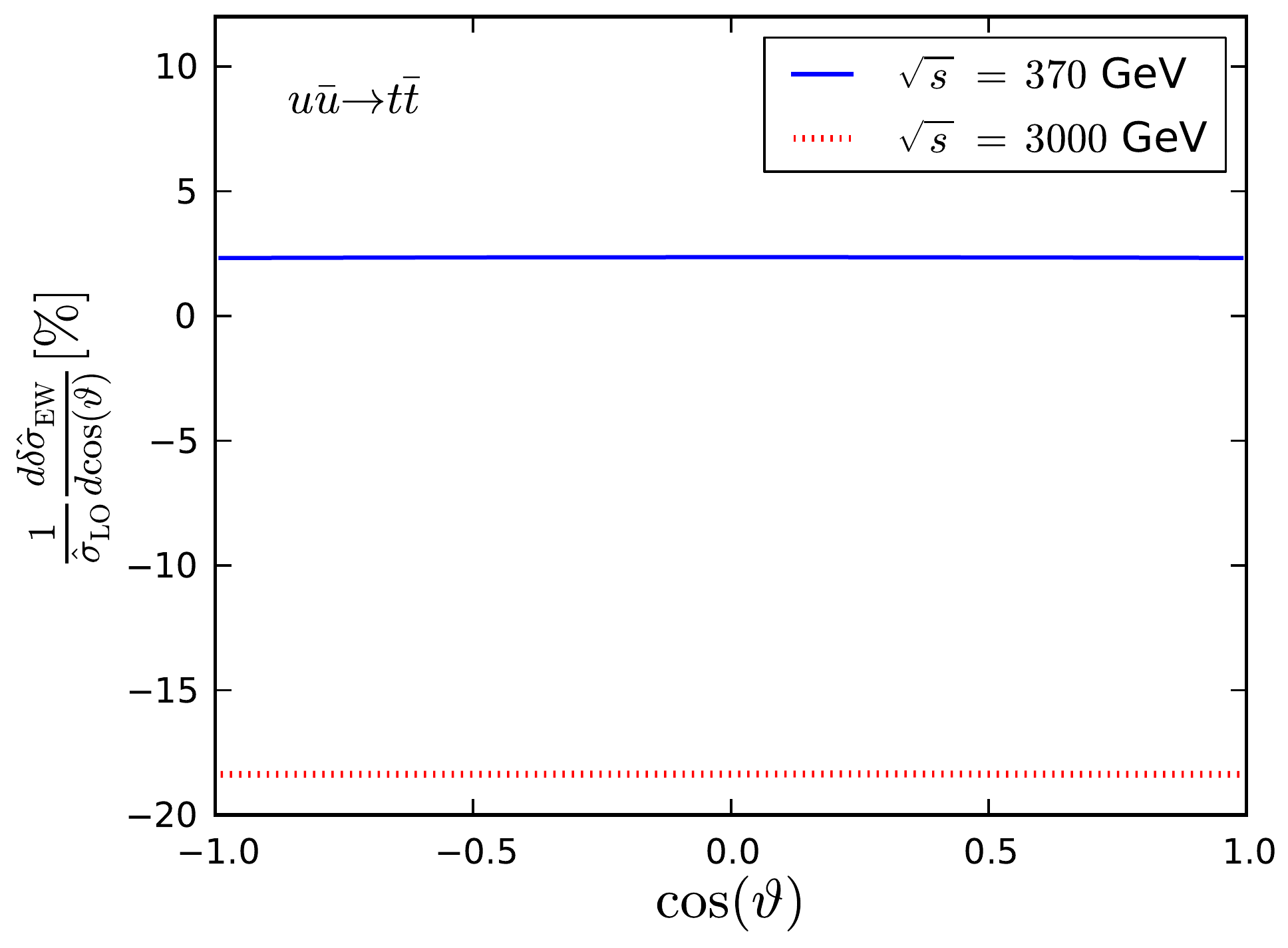}
    \includegraphics[width=0.45\textwidth]{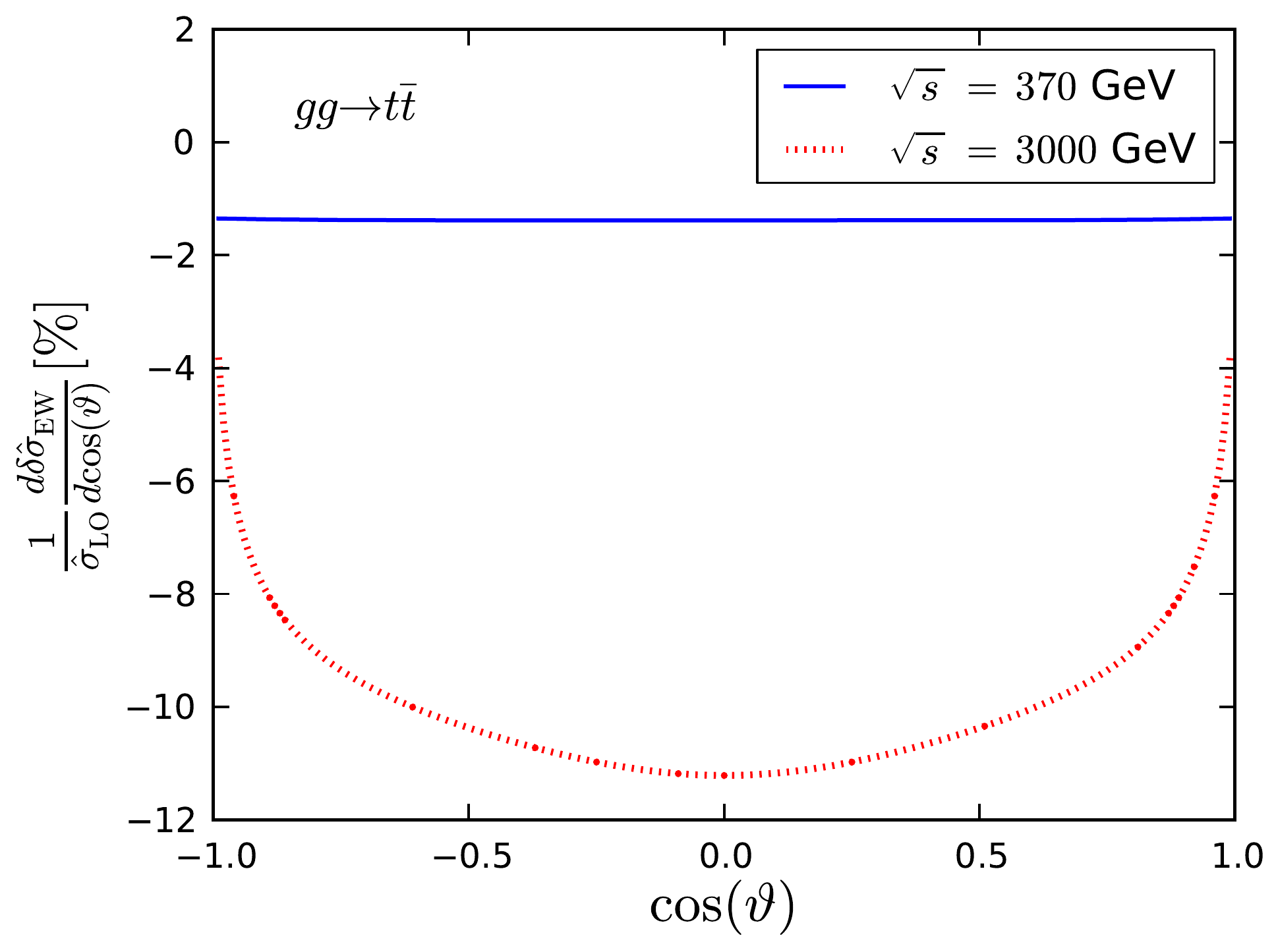}
    \caption{\label{fig:angularcor}%
      Relative weak corrections for the quark-  and gluon-induced
      reactions as functions of the scattering angle close to threshold
      ($\sqrt{\hat s}=370\, {\rm GeV}$, solid line) and at high 
      energies ($\sqrt{\hat s}=3\,{\rm TeV}$, dotted line)}
  \end{center}
\end{figure}   

Let us now enter the description of the corrections in more detail.
The corrections at the partonic level are shown in
Fig.~\ref{fig:partoncor} for quark and gluon induced processes as
functions of $\hat s$. For the quark--anti-quark channel we include
only the infrared finite vertex corrections which are responsible for
the Sudakow suppression at large momentum transfer. The box
contributions for the $q\bar q$ process are important only for the
charge-asymmetric piece
\cite{Kuhn:2011ri,Hollik:2011ps,Bernreuther:2012sx,%
  Ahrens:2011uf,Almeida:2008ug}, and can be neglected in the present
context.  As expected, away from very small $\hat s$ the corrections
are negative and about twice as large for quark- compared to
gluon-induced processes. Only very close to threshold one observes
corrections which become positive for a light Higgs boson and will be
discussed in section 3. For the ficticious case of $M_H=1~{\rm TeV}$
two pronounced structures are visible in the gluon-fusion channel: The
interference between the Born amplitude and the $s$-channel Higgs
boson contribution (last diagram of Fig.3) is visible as slight
depletion around 1~TeV, the interference with the $Z$ plus $\chi$
contribution arising from the same diagram is responsible for the dip
close to threshold. For $M_H=126~{\rm GeV}$ this dip is
overcompensated by the positive contribution of roughly 5\% from the
Yukawa interaction discussed in more detail in section 3. This same
difference of 5\% between $M_H= 126~{\rm GeV}$ and $1~{\rm TeV}$ is
also visible in the threshold behaviour of the $q\bar q$-initiated
reaction. 

The angular dependence of the corrections is shown in
Fig.\,\ref{fig:angularcor} separately for quark and gluon induced
processes close to threshold at 370 GeV (upper solid blue curve) and for 3
TeV (lower dotted red line). 
Let us, in a first step, discuss the results for the quark-induced
reaction (Fig.~\ref{fig:angularcor} left).
Again we restrict the analysis to the vertex
correction. Close to threshold the process is dominated by (isotropic)
$S$-waves, at high energies ($3~{\rm TeV}$) the Dirac form factor
dominates both for Born amplitude and correction.
This leads to a constant ratio as function of the scattering angle. At
low energies ($\sqrt{\hat s}=370$~GeV) we find a positive correction
of about 2 \%. At large energies, say $\sqrt{\hat s}=3{\rm TeV}$,
the Sudakow suppression leads to
negative corrections of about $-18$ \%. Note that the box diagrams
while not particularly enhanced would lead to sizable
asymmetric and small symmetric corrections. 
For details we refer to Ref.~\cite{Hollik:2011ps}. 
The gluon induced part, in contrast, is markedly angular dependent.
For large $\hat s$ and small scattering angle the corrections are
small, since the Sudakov-like behaviour cannot be expected in this
case. At ninety degrees, in contrast, the Sudakov limit is applicable
and the corrections become large.
\begin{figure}
  \begin{center}
    \includegraphics[width=0.6\textwidth]{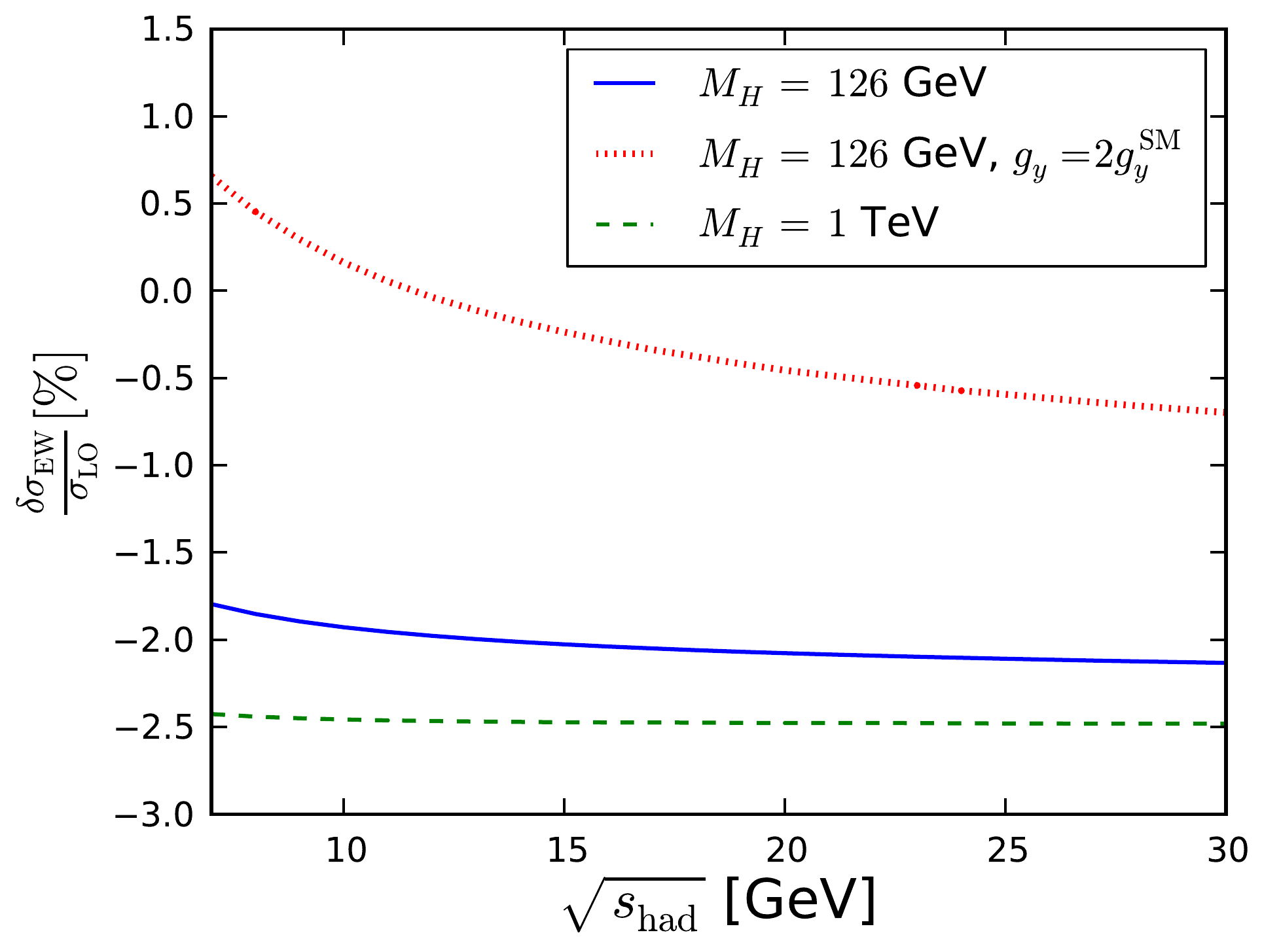}
    \caption{\label{fig:sigtot}%
      Relative weak corrections for the total cross section as
      functions of the total cms energy for two different masses of the
      Higgs boson and for a rescaled Yukawa coupling.}
  \end{center}
\end{figure}   
\begin{figure}
  \begin{center}
    \includegraphics[width=0.48\textwidth]{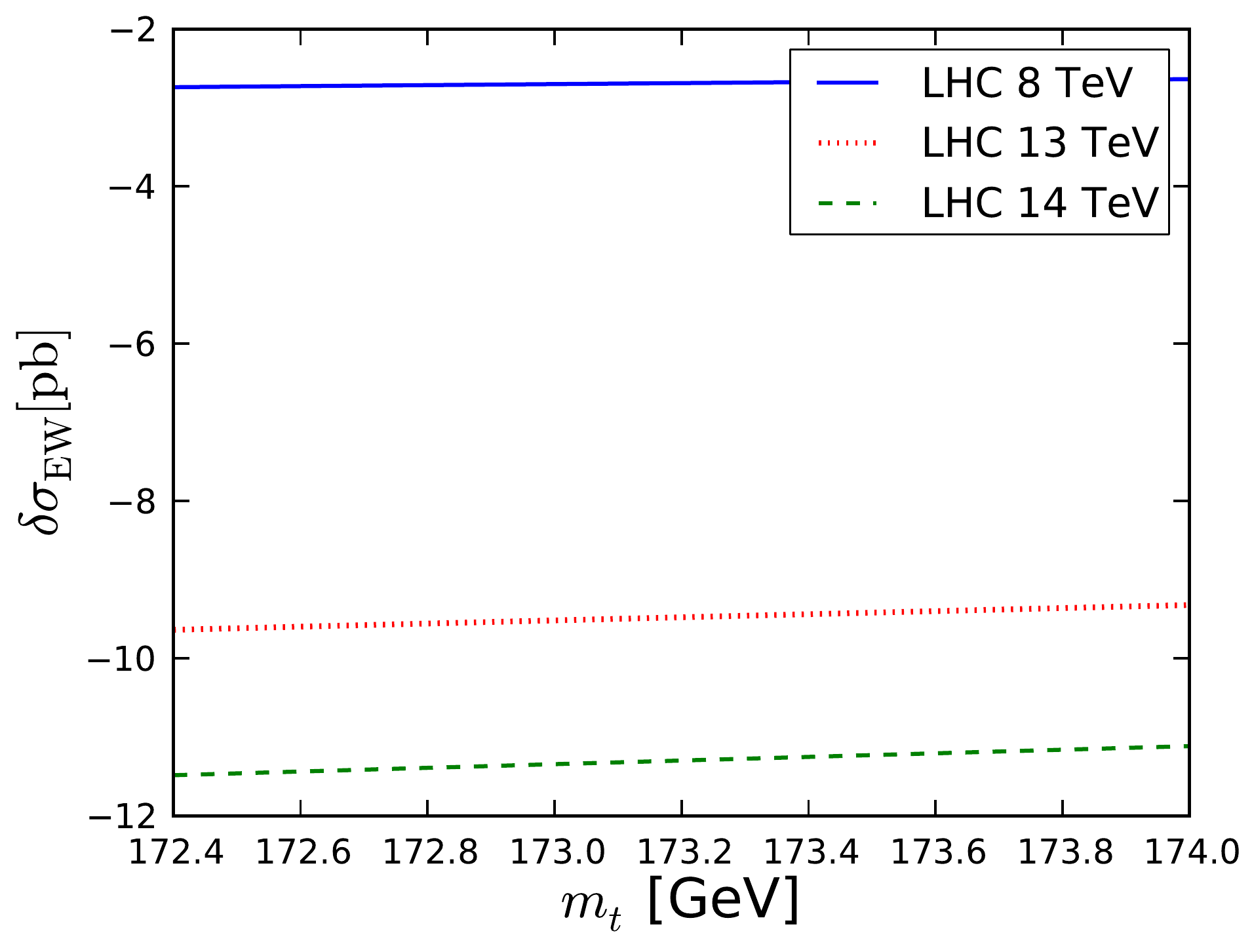}
    \includegraphics[width=0.48\textwidth]{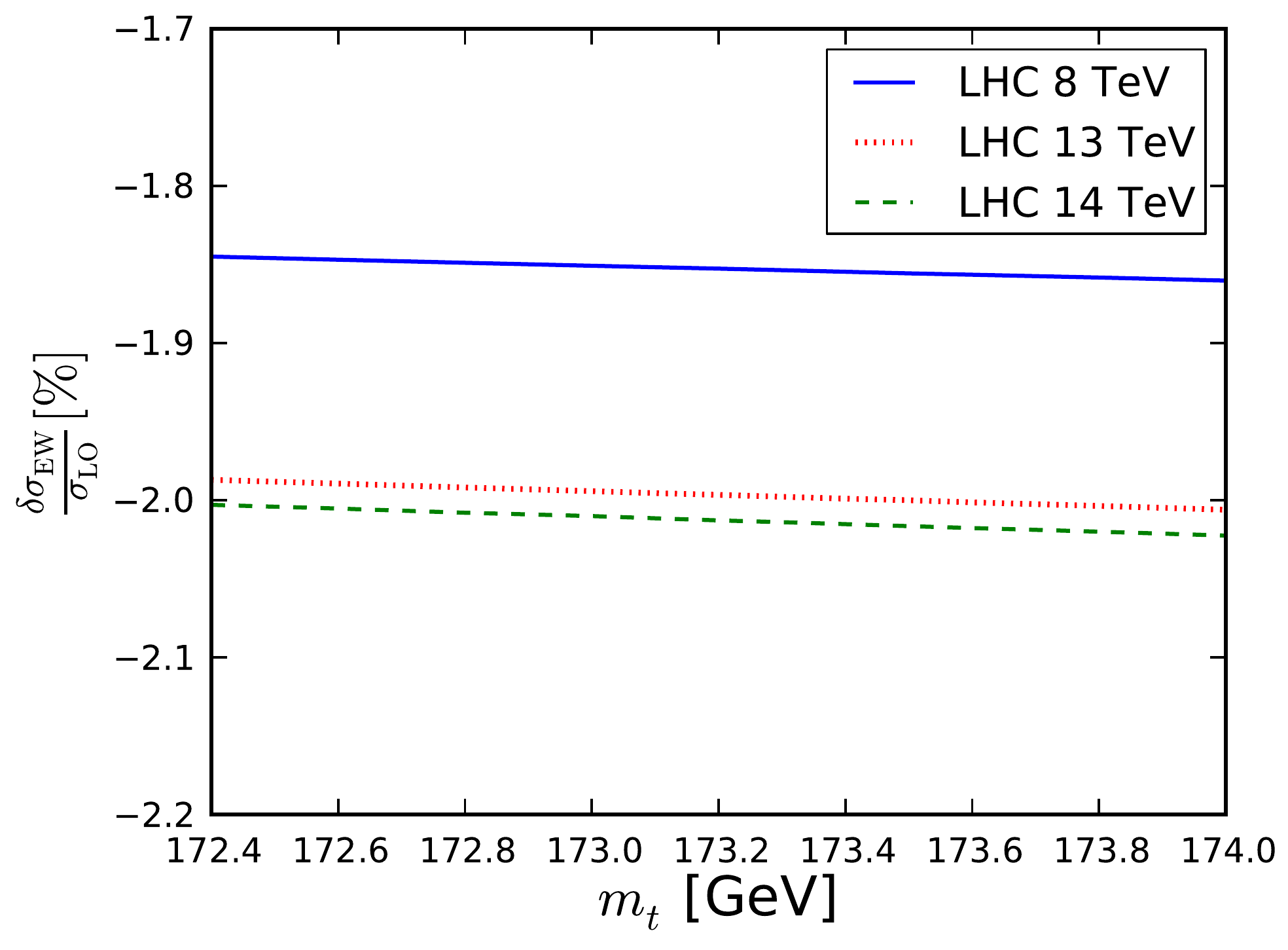}
    \caption{\label{fig:TopMassDependence}
      Weak corrections as function of the top-quark mass.}
     \end{center}
\end{figure}   
Let us now discuss observables at the hadron level. In difference to
the discussion at parton level we include in the analysis now also
box-contributions and real corrections, thus the full set of
corrections are investigated.
The corrections for the total cross section are shown in
Fig.~\ref{fig:sigtot} as function of $\sqrt{s}$, for two
characteristic choices of the Higgs mass, $M_H=126~{\rm GeV}$ and
$1000~{\rm GeV}$. We allow $M_H$ to move away from its recently
determined value to illustrate the effect of
the Higgs-top Yukawa coupling. The corrections are evidently small, of
order minus two percent for all LHC energies and only moderately
sensitive to $M_H$.  Given the recent progress concerning the NNLO QCD
calculations the theoretical uncertainties will eventually reach
3--4\%.  At this level of accuracy the weak corrections become
important and need to be taken into account. As reference we show in
Fig.~\ref{fig:TopMassDependence} the weak correction as
function of the top-quark mass. At 8 TeV centre of mass energy the
corrections are about $-1.85$\%. At 14 TeV the high-energy regime of
the cross section becomes more accessible leading to slightly larger
corrections of the order of $-2.0$\%.
For convenience we provide a parametrisation for the weak corrections
valid for $172.4 < M_t < 174$: 
\def\pb{\mbox{pb}}
\begin{eqnarray}
  \delta\sigma^{\rm 8TeV}_{\EW}  &=& -2.69 \pb + 0.06\pb \times (M_t/\GeV-173.2) \\
  \delta\sigma^{\rm 13TeV}_{\EW} &=& -9.48 \pb + 0.20\pb \times (M_t/\GeV-173.2) \\
  \delta\sigma^{\rm 14TeV}_{\EW} &=& -11.30 \pb + 0.23\pb \times (M_t/\GeV-173.2)
\end{eqnarray}
As can be seen from Fig.~\ref{fig:TopMassDependence} and the
parametrization above, the mass dependence is very small in the range
172.4 -- 174 GeV and can be neglected for most phenomenological applications.
The aforementioned results can be directly combined with NNLO results
calculated using the MSTW2008NNLO PDF set \cite{Martin:2009iq}. 
Since ratios are more robust with respect to different choices for the
parton distribution functions we also present a parametrization for
the relative corrections:
\begin{eqnarray}
  {\delta\sigma^{\rm 8TeV}_{\EW}\over \sigma_{\LO}} 
  = -1.85 \%   -0.01\%\, \times  (M_t/\mbox{GeV}-173.2), \\
  {\delta\sigma^{\rm 13TeV}_{\EW}\over \sigma_{\LO}} = 
  -2.00 \%   -0.01\%\,\times  (M_t/\mbox{GeV}-173.2), \\
  {\delta\sigma^{\rm 14TeV}_{\EW}\over \sigma_{\LO}} = 
  -2.01\%   -0.01\%\,\times  (M_t/\mbox{GeV}-173.2). 
\end{eqnarray}
Note that the coupling constants of the strong interactions cancels
in the ratios. In addition, for the ratios the mass dependence is
further reduced and completely neglible in the range 172.4 -- 174 GeV.
One may argue that some contributions of the QCD corrections are
universal and will also correct the weak contributions (see also the
discussion in section \ref{sec:QCD-EW-combined}). Based on this
assumption one may use
\begin{equation}
  \delta \sigma_{\EW} =  
  {\delta\sigma_{\EW}\over \sigma_{\LO}} \times \sigma_{\NNLO}^{\QCD}
\end{equation}
to estimate the size of the weak corrections.
For $M_t= 173.2$ GeV this leads to
\begin{eqnarray}
  \delta\sigma^{\rm 8TeV}_{\EW}  &=& -4.4\pb,\\
  \delta\sigma^{\rm 13TeV}_{\EW} &=& -15.8\pb, \\
  \delta\sigma^{\rm 14TeV}_{\EW} &=& -18.8\pb.
\end{eqnarray}
\begin{figure}
  \begin{center}
    \includegraphics[width=0.5\textwidth]{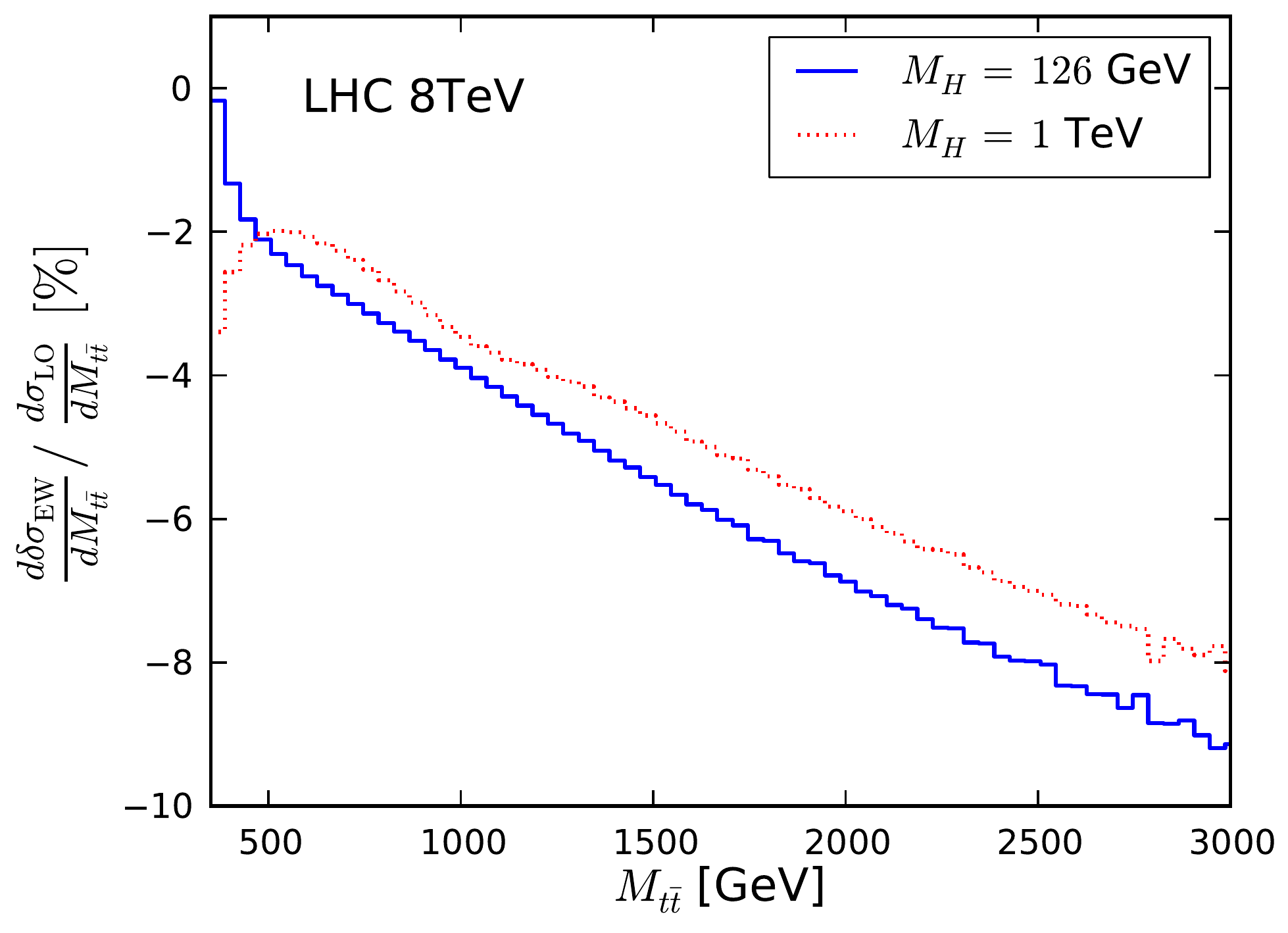}
    \includegraphics[width=0.48\textwidth]{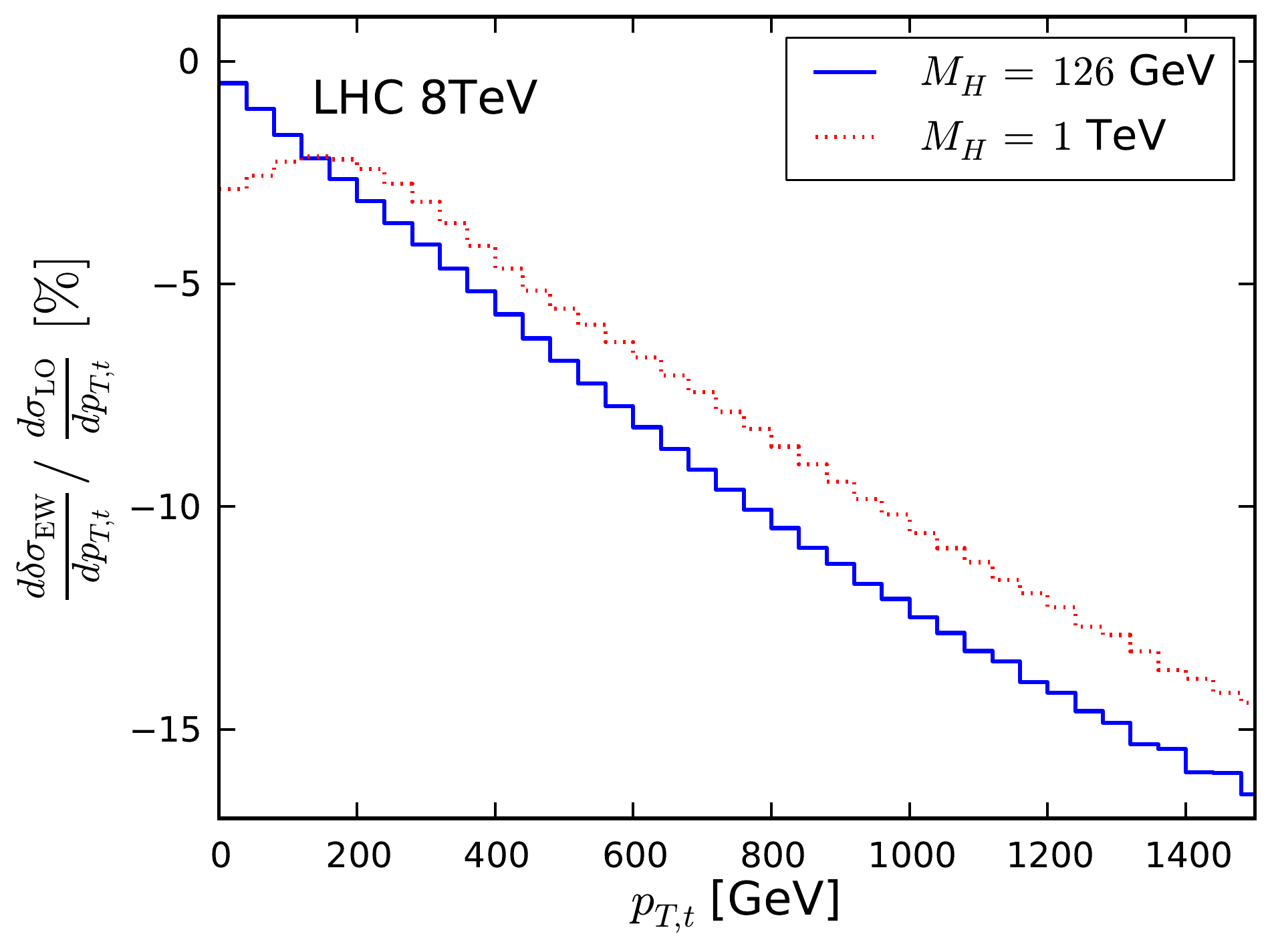}
 \caption{\label{fig:distr8}%
Relative weak corrections for the invariant $t\bar t$ mass (left) and 
transverse momentum (right) distribution for LHC 
for Higgs masses of 126~GeV and 1~TeV. 
}
     \end{center}
\end{figure}
In addition we demonstrate in Fig.~\ref{fig:sigtot} the impact of an
enhanced Yukawa coupling with $g_Y=2g_Y^{SM}$.
In this case the negative corrections from the large transverse
momentum region are overcompensated by the positive ones for small
$t\bar t$ masses.
Let us emphasize that such an analysis might well lead to a non-trivial
limit on the top quark Yukawa coupling $g_Y$. The weak
corrections have a trivial dependence on the Yukawa coupling. A large
fraction of the corrections does not depend on $g_Y$. The production
of an intermediate $s$-channel Higgs boson through a closed $b$-quark
loop leads to a contribution linear in the top-quark Yukawa
coupling. Due to the small b-quark mass, this contribution is expected
to be very small. The remaining dependence on the top-quark Yukawa
coupling is quadratic, since it allways involves two $Ht\bar t$ vertices. 
For the total cross section we find the following parametric
dependence ($M_t=173.2$ GeV):
\begin{eqnarray}
  {\delta\sigma^{\rm  8 TeV}_{\rm EW}} &=& 
  ( -3.80  + 0.0009 g_Y + 1.12 g_Y^2 ) \pb, \\
  {\delta\sigma^{\rm 13 TeV}_{\rm EW}} &=& 
  ( -12.47  + 0.0136 g_Y + 2.99 g_Y^2 ) \pb, \\
  {\delta\sigma^{\rm 14 TeV}_{\rm EW}} &=& 
  ( -14.74  + 0.0146 g_Y + 3.45 g_Y^2 ) \pb.
\end{eqnarray}
Indeed we observe that the linear dependence on $g_Y$ is very weak and
can be neglected as anticipated above. 
Again it might be useful to study the size of the relative corrections:
\begin{eqnarray}
  {\delta\sigma^{\rm  8 TeV}_{\rm EW}\over\sigma_{\LO} } &=& 
  ( -2.62  + 0.0006 g_Y + 0.77 g_Y^2 ) \%, \\
  {\delta\sigma^{\rm 13 TeV}_{\rm EW}\over\sigma_{\LO} } &=& 
  ( -2.63  + 0.0029 g_Y + 0.63 g_Y^2 ) \%, \\
  {\delta\sigma^{\rm 14 TeV}_{\rm EW}\over\sigma_{\LO} } &=& 
  ( -2.63  + 0.0026 g_Y + 0.61 g_Y^2 ) \%.
\end{eqnarray}
A shift of $g_Y$ by a factor three would enhance the cross section by
about 5\% and might thus become experimentally accessible. As
discussed in section \ref{sec-threshold}, this limit can be further
improved by restricting the sample to events close to the production
threshold.

While for the total cross section the weak corrections are small, the
situation is drastically different, once we consider differential
distributions in the region of large transverse momenta $\pt$ or large
masses $M_{t\bar t}$ of the $t\bar t$ system where Sudakov logarithms
start to play an important role.  The corrections are shown in
Fig.~\ref{fig:distr8} for proton-proton collisions with center of mass
energies of 8~TeV and 14~TeV both for the $\pt$- and the $M_{t\bar
  t}$-distributions. (Results for 13 and 14 TeV are shown in the
Appendix.) For illustration we present again the relative
corrections for Higgs masses of 126~GeV and 1~TeV. The strong increase
with increasing $\pt$ is evident.  Based on the present data sample,
corresponding to to more than 20~${\rm fb}^{-1}$, corrections close to
-10\% could be observed at $\sqrt{s}=8~{\rm TeV}$ if events with top
quarks of large transverse momenta, say 750~GeV, are considered.

\begin{figure}
  \begin{center}
    \includegraphics[width=0.45\textwidth]{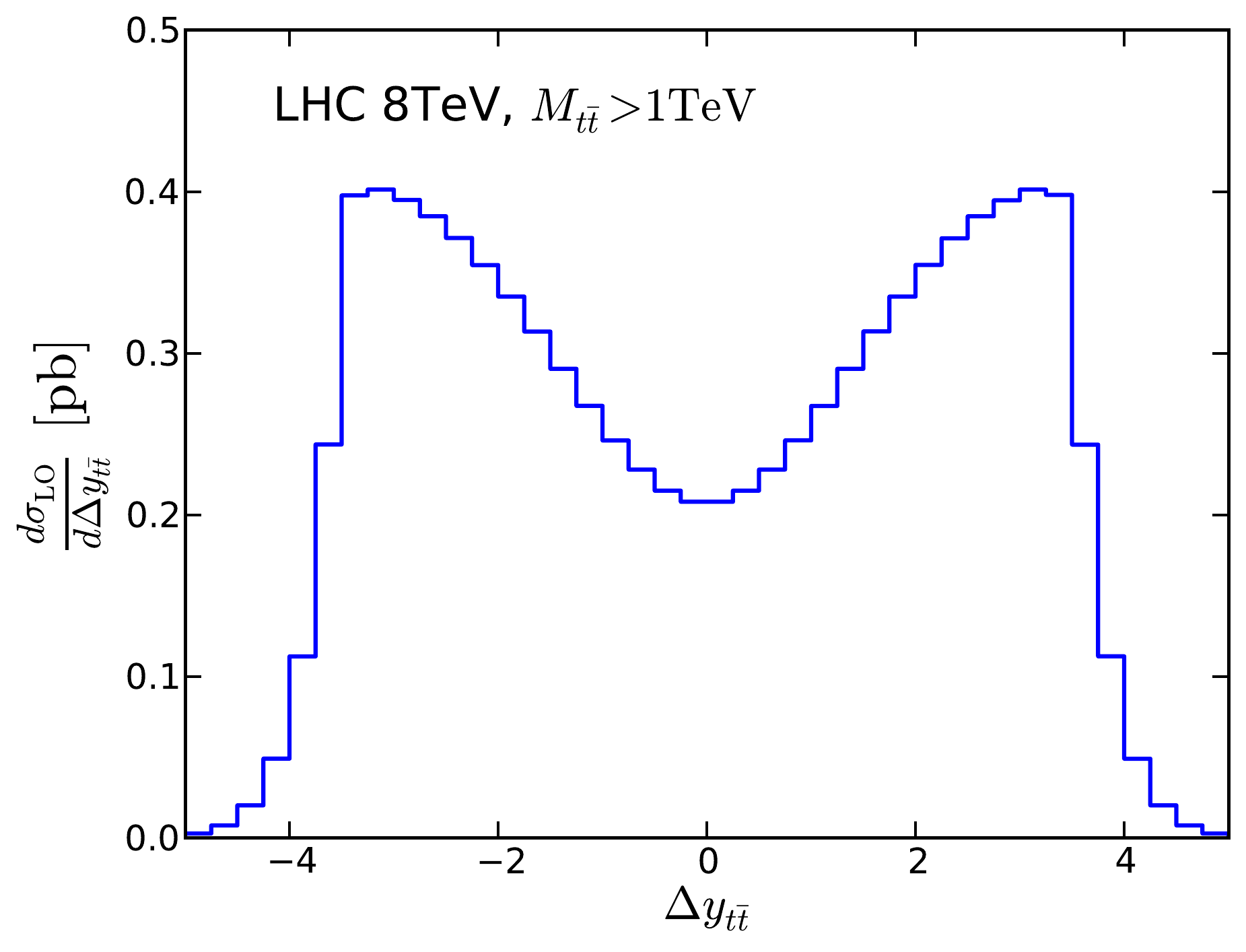}
    \includegraphics[width=0.45\textwidth]{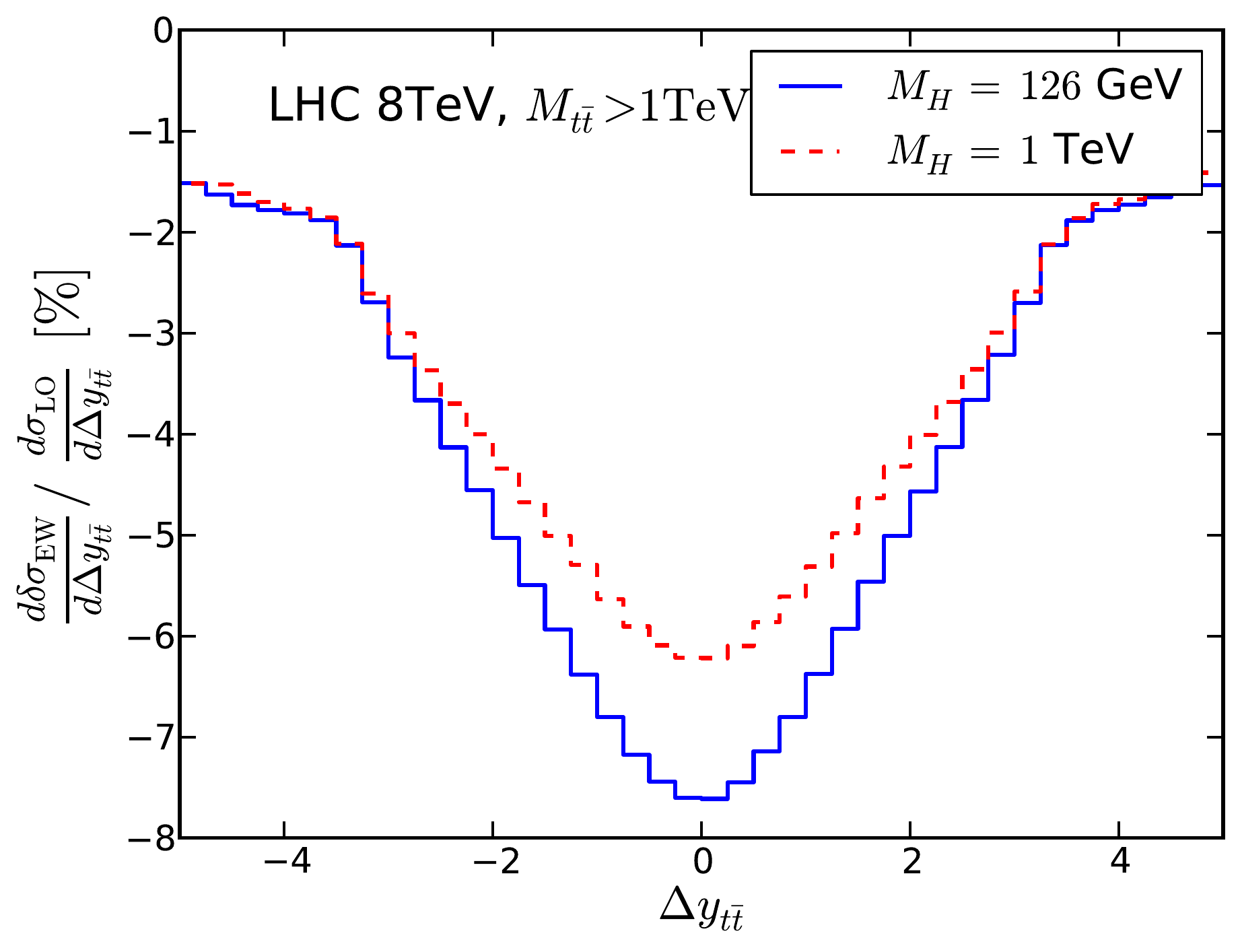}
 \caption{\label{fig:rapidity-distr}%
Rapidity distributions with invariant mass cuts at leading order (left) and relative weak corrections (right)}
     \end{center}
\end{figure}

To investigate the angular dependence of the $t\bar t$ system in its
center of mass frame one could consider the distribution in the
rapidity difference $\Delta y_{t\bar t}=y_{t}-y_{\bar t} $ which, for
fixed $M_{t\bar t}$ can be directly translated into the angular
distribution.  To illustrate the distributions and the size of the
corrections, the differential distributions $d\sigma/d\Delta y_{t\bar
  t}$ are shown in Fig.\ref{fig:rapidity-distr} for 8~TeV, considering
only events with $M_{t\bar t}$ larger than 1~TeV. (For 13 and 14 TeV
the results are shown in the Appendix.)  The corresponding relative
corrections are also displayed in Fig.\ref{fig:rapidity-distr}.  The
pronounced peaking of the cross section for large rapidity differences
in Fig.\ref{fig:rapidity-distr} (left) is an obvious consequence of
the $t$-channel singularity, the enhanced negative corrections around
$\Delta y_{t\bar t}= 0$ in Fig.\ref{fig:rapidity-distr} (right) are a
consequence of the Sudakov condition $\hat s$ and $ |{\hat t}|\gg
M_W^2$. Since the distribution in $\Delta y_{t\bar t}$ is at the same
time sensitive to anomalous couplings, these could well be masked by
the large radiative corrections.

\section{QCD and electroweak corrections combined}
\label{sec:QCD-EW-combined}
Let us at this point speculate about the combination of weak and
QCD corrections. Clearly, the evaluation of corrections of ${\cal
O}(\alpha_s\alpha)$ is out of reach in the foreseeable future. 
Thus, strictly speaking, both a multiplicative 
(of the form $(1+\delta_{QCD})(1+\delta_{W}))$ and an additive
(of the form $(1+\delta_{QCD}+\delta_{W}))$ treatment is equally
justified. The difference between the two assumptions can be
considered as an estimate of the theory uncertainty. It may be
usefull, however, to devise a strategy to implement eventually 
the major part of the combined corrections.
As mentioned in the beginning, QED and purely weak corrections can be
treated seperately in the present case. Furthermore, QED corrections are
small and the resulting uncertainty of combined QCD and QED terms even
smaller. In principle, by adjusting color coefficients, the recently
available two-loop QCD corrections 
\cite{Czakon:2012zr,Czakon:2012pz,Czakon:2013goa}
could be employed to arrive at the
full combined QED and QCD results. Concerning the weak corrections, we
observe that a major part of the QCD corrections originates from
configurations involving soft and/or collinear emission. Let us then 
reconstruct the effective two-body kinematics by using the $t\bar t$ 
invariant mass as $\hat s$ and the scattering angle with respect to 
the beam direction, as defined in the $t\bar t$ rest frame as 
partonic scattering angle. Using this information would allow to apply
the weak correction factor which also depends on $\hat s$ 
and $\hat t$ only. 

Let us describe the prescriptions in more detail. As discussed before,
the corrections depicted in Figs.\ref{fig:RealCor} and
\ref{fig:IR-cancellation} can be ignored. Three overall correction
factors $K_{u\bar u}$, $K_{d\bar d}$ and $K_{gg}$ 
with
\begin{displaymath}
  K_{ij} = {d\sigma_{ij}^{\EW}/d\cos(\theta)\over d\sigma_{ij}^{\LO}/d\cos(\theta)}
\end{displaymath}
remain, which are for
given $M_H$, $M_W$ and $M_Z$ functions of $\hat s$ and $\hat t$ and
are appropriate for the three basic two-to-two processes.  Numerical
results for the $K_{u\bar u}$ and $K_{gg}$ are shown in
Fig.~\ref{fig:angularcor} with $\sqrt{\hat s}$ fixed to 370, and
3000~GeV.  We have implemented the corresponding analytic formulae in
the publicly available Hathor program \cite{Aliev:2010zk,Kant:2014oha}
version 2.1 which is available at {\tt
  http://www.physik.hu-berlin.de/pep/tools/hathor.html}.
The correction factors for $u$ and $d$ quarks are nearly the
same and angular independent, the correction factor for the
gluon-induced process is most pronounced for top quarks produced in
the transverse direction.

It is important to keep in mind that the corrections do not affect the
two-to-two kinematics. This allows to implement the electroweak
corrections into any Monte Carlo generator for $t\bar t$ production as
follows: In a first step we consider a
generator which does not involve NLO QCD corrections.  The invariant
mass of the $t\bar t$ system will be identified with $\sqrt{\hat s}$.
Let us now denote the directions of the momenta of top and antitop
quark in the $t\bar t$ rest frame by $\bf \vec{e}^*_t$ and $\bf
\vec{e}^*_{\bar t}$, respectively, the directions of the beam
momenta by $\bf \vec{e}^*_1$ and $\bf \vec{e}^*_2$, the direction of
the effective scattering axis by\footnote{This approach follows the
  one introduced in Ref.~\cite{Gieseke:2014gka} for gauge-boson pair
  production.}
\begin{equation}
  {\bf \hat{\vec{e}}^*}
  \equiv 
  \frac{{\bf \vec{e}^*_1} - {\bf \vec{e}^*_2}}
  {|{\bf \vec{e}^*_1} - {\bf \vec{e}^*_2}|}
\end{equation}
and the effective scattering angle by
\begin{equation}
\cos\theta^*\equiv {\bf \vec{e}^*_t} \cdot {\bf \hat{\vec{e}}^*}.
\end{equation}

The partonic variables are thus obtained from
\begin{equation}
\label{eq:kinematics}
\hat s \equiv M_{t\bar t}^2, \,\,\, 
\hat t\equiv m_t^2-\frac{\hat s}{2}(1-\sqrt{1-\frac{4 m_t^2}{\hat s}}\cos\theta^*).
\end{equation}

Inspecting the event, as generated through the program, more closely, 
it can be assigned in a unique way to the $u\bar u$, $d\bar d$ or $gg$
induced subprocess. Using the correction functions $K_i(\hat s, \hat t)$ 
with $i=u\bar u$, $d\bar d$ or $gg$, the reweighting can be performed in
a straightforward way.

The kinematic prescription outlined above can also be applied to
generators which include NLO QCD corrections. Events which involve
collinear or soft quark or gluon emission are responsible for the
major part of QCD corrections. For these events one would expect that
the dominant weak corrections can still be derived from the same
correction functions with $\hat s$ and $\hat t$ as derived from
eq.~(\ref{eq:kinematics}) above.  For final states with the $t\bar t$
system produced at large transverse momentum, balanced by a hard quark
or gluon jet at large $P_t$, our prescription will no longer properly
account for the electroweak corrections. However, the missing terms
are of order $\alpha_s\alpha_{weak}$ and their evaluation would
require the electroweak corrections for the full set of two-to-three
reactions.

It remains to assign the proper correction function $K_i$ for the full
set of partonic processes. For the subprocesses
$q\bar q\to t\bar t (g)$ and $gg\to t\bar t(g)$ the functions 
$K_{q\bar q}$ and $K_{gg}$ will be employed. The assignment of
correction factors to the reaction $qg\to t\bar t q$ (and its charge
conjugate) is more involved. Let us assume that the incoming quark
originates from hadron 1 and splits into a nearly collinear quark and
gluon. The latter fuses with the gluon from hadron 2 into $t\bar t$. In
this case the factor $K_{gg}$ is suggested. Alternatively one may
consider a situation where the incoming gluon splits into a nearly
collinear $q\bar q$ pair, with the anti-quark annihilating into 
$t\bar t$. In this case the use of $K_{q\bar q}$ seems more adequate.

To distinguish the two options, we consider the scattering angle
$\theta_q$ of the outgoing quark relative to hadron 1 in the $t\bar t$
rest frame.
If $0\le  \theta_q<\pi/2$, we take $K_{gg}$, if $\pi/2 \le
\theta_q<\pi$,
we take $K_{q\bar q}$. In the limiting cases of nearly collinear
emission this convention leads to the desired result, in the
case of events with a quark jet at large transverse momentum 
the expected error will be of order $\alpha_s\alpha_{weak}$. 

An alternative method to define a reduced kinematic in case of
additional emission could be using a boost similar to what has been
done in the matrix element method. The sensitivity to the specific
prescription could be estimated by comparing the two different approaches.

\section{The top-pair threshold region and the Yukawa coupling}
\label{sec-threshold}

As illustrated in Fig.~\ref{fig:partoncor}, the corrections for
top-pair production very close to threshold exhibit a significant
dependence on the mass of the Higgs boson.  In fact, both for quark
and gluon induced process the difference in the correction between a
light ($M_H=126~{\rm GeV}$) and a heavy ($M_H=1000~{\rm GeV}$) Higgs
boson amounts to about 5\%. This effect has been discussed in some
detail for pair production at an electron-positron collider
\cite{Grzadkowski:1986pm,Strassler:1990nw,Fadin:1990nz,Jezabek:1993tj,Harlander:1995dp}
and for quark-antiquark collisions \cite{Harlander:1995dp} and is
closely related to the well-known Sommerfeld rescattering corrections,
originally obtained in the framework of QED.  Similar considerations
are also applicable to gluon fusion \cite{Kuhn:2006vh}.

For a Yukawa potential induced by the Higgs exchange,
\be
V_Y(r)= - \kappa\; {1\over r}\,e^{-r/r_Y}\,\,{\rm with}\,\,
\kappa=\frac{g_Y^2}{4\pi}=\frac{\sqrt{2} G_F M^2_t}{4\pi} \approx 0.0337
\,\,{\rm and}\,\, r_Y=1/M_H,
\ee
the dominant correction evaluated directly at threshold 
is given by the factor $1+\kappa\frac{M_t}{M_H}$.
(The full result including the energy dependence,
can be found in \cite{Grzadkowski:1986pm,Jezabek:1993tj}.) 
Indeed the difference of 5\% 
between the heavy and the light Higgs boson is well
consistent with this simple approximation. For quark-antiquark annihilation
the positive offset is shown in Fig.\ref{fig:partoncor} (left).
For gluon fusion the Yukawa
enhancement is partially masked by a negative contribution originating 
from the interference of the tree-level amplitude with the amplitude from
the triangle diagrams with $Z$ and $\chi$ in the $s$-channel
(Fig.~\ref{fig:VirtualCor-gg}). 
The difference, however, between a heavy and a light Higgs boson
of about 5\% remains unchanged.

As evident from Fig.\ref{fig:partoncor}, the Yukawa enhancement is located in
the region close to threshold, with relative $t\bar t$-velocity 
$\beta$ less than $M_H/M_t$.
For the moment we consider the weak corrections as an overall
$\beta$-dependent factor which multiplies the complicated threshold
behaviour induced by the partly attractive, partly repulsive QCD
potential. (For QCD effects see e.g. \cite{Kiyo:2008bv} and 
references therein.) 
In principle the effect of a light Higgs exchange could be split into a
short range piece, which leads to a  $\beta$-independent correction
term, and a long-range piece, which can be absorbed by adding 
Yukawa and QCD potential. The energy dependence can then be obtained from a
Green's function treatment. This approach has been discussed in more
detail in \cite{Harlander:1995dp} for the cases of top production in 
electron-positron
and quark-antiquark annihilation. Since $r_Y$,
the characteristic lenght of the Yukawa potential, is still 
significantly smaller than $r_B$, the Bohr radius of the would-be toponium
ground state, 
\be
r_Y/r_B=(\frac{4}{3}\alpha_s \frac{M_t}{2})/M_H\approx 1/6,
\ee
the simple multiplicative treatment advocated above is sufficient for
the presently required level of precision.

\begin{figure}
  \begin{center}
    \includegraphics[width=0.6\textwidth]{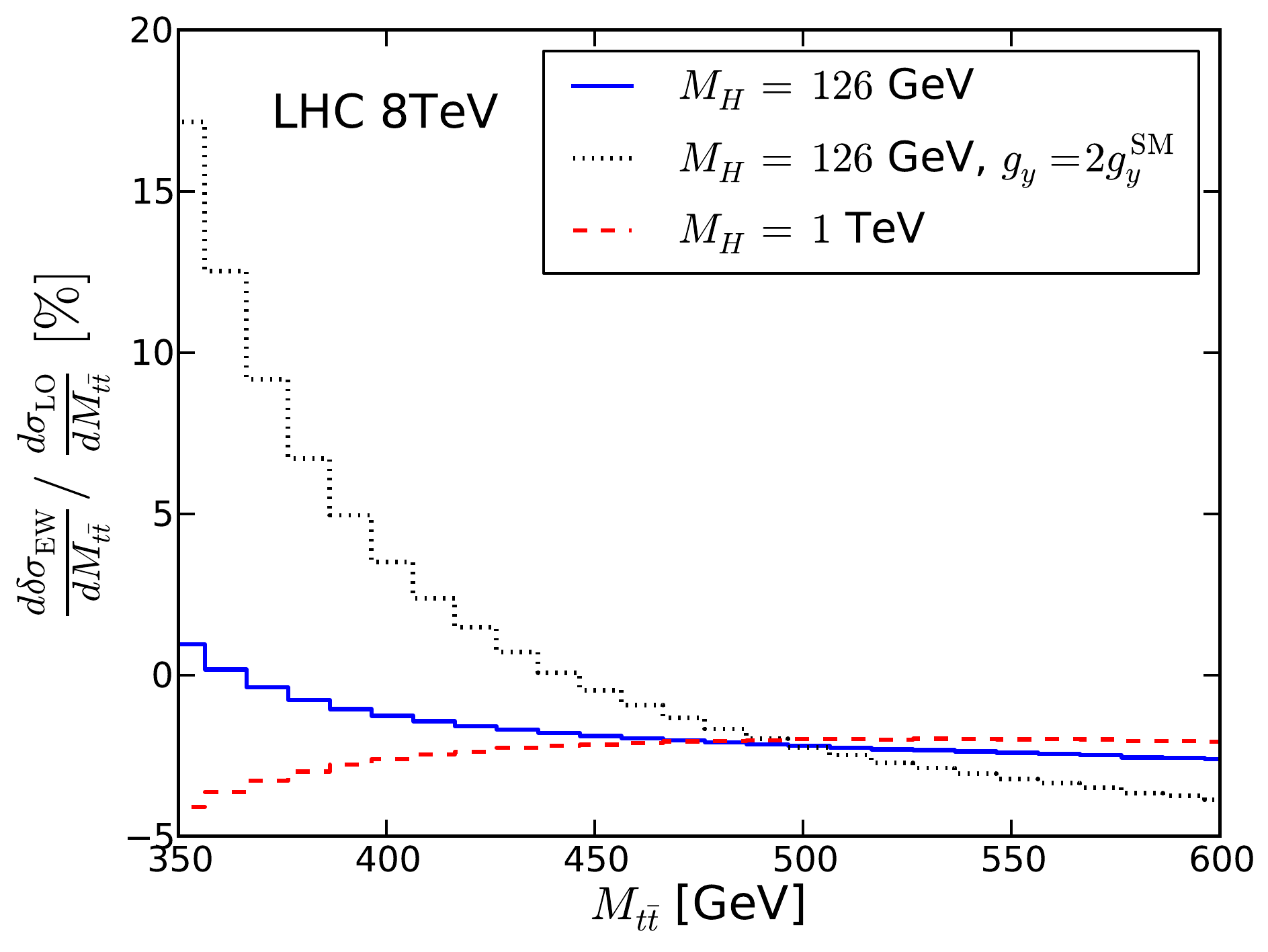}
    \caption{\label{fig:threshold-distr}%
      Relative weak corrections for the mass distribution in the
      framework of the SM assuming  $M_H=126~{\rm GeV}$ (solid blue curve) 
      and 1000~GeV (dashed red curve), and for the case of an enhanced Yukawa
      coupling   $g_Y=2 g^{SM}_Y$ with  $M_H=126~{\rm GeV}$ 
      (dotted black curve).  
      }
  \end{center}
\end{figure}

\begin{figure}
  \begin{center}
    \includegraphics[width=0.6\textwidth]{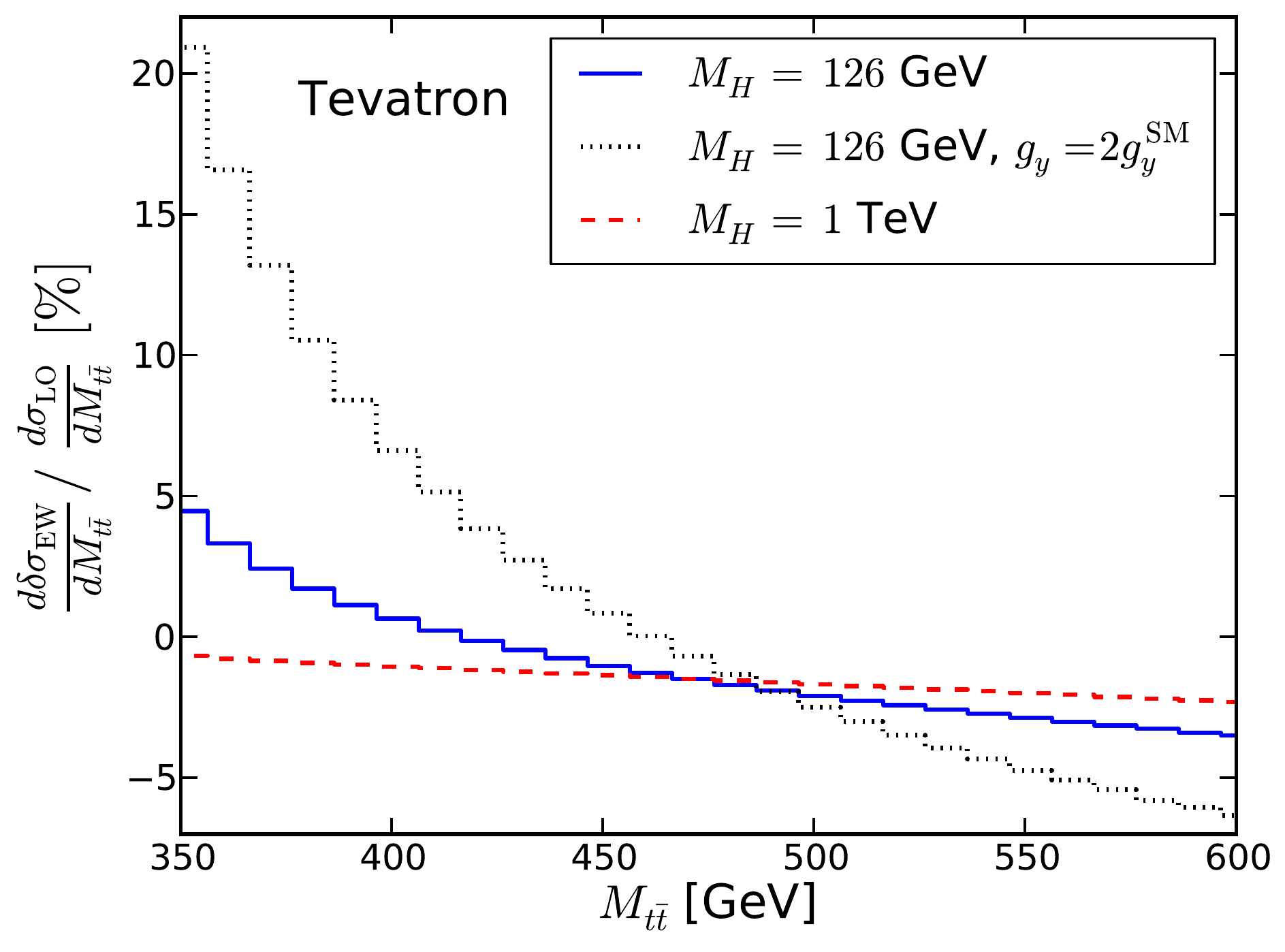}

 \caption{\label{fig:threshold-distr-tev}%
Same as Fig.\ref{fig:threshold-distr} but for the Tevatron.
}
     \end{center}
\end{figure}

As discussed above, the impact on the total cross section from the
variation of $M_H$ is relatively small, less than one percent, both
for the Tevatron and the LHC, and even an enhancement of the Yukawa
coupling by a factor two will be hardly visible.  Differential
distributions, however, are significantly more sensitive to the Yukawa
coupling. This is demonstrated in Figs.~\ref{fig:threshold-distr},
\ref{fig:threshold-distr-tev}, where the correction factors for the
distribution with respect to $M_{t\bar t}$ are evaluated for the LHC
at 8 TeV and Tevatron in the region close to threshold. (Results for
13 and 14 TeV are given in the Appendix.)
 
As expected from the previous discussion, differences 
around 5\% between the cases $M_H= 126~{\rm GeV}$ and 1~TeV are visible.
It remains to be seen, whether the experimental mass resolution and
normalization of the cross section will be sufficiently precise to pin
down the 5\%-effect and thus determine directly the Yukawa coupling
$g_Y$.  At the same time this approach requires a detailed theoretical
understanding of the QCD predictions for the threshold behaviour,
governed by the remnants of the bound states, as discussed in
\cite{Kiyo:2008bv}.  However, in any case this approach should allow
to provide an upper limit on modifications of $g_Y$ that might be
postulated in theories beyond the Standard Model 
\footnote{For examples where this has been discussed in the framework of
  a two-Higgs-doublet model see e.g.  Refs.\cite{Denner:1991tb,Guth:1991ab}}

Let us assume, for example, the case of an enhanced Yukawa coupling
$g_Y=2g_Y^{SM}$. This magnifies the Yukawa correction by a factor four
and implies an enhancement of the cross section close to threshold by
about 20\%. (See dashed curves in
Figs.~\ref{fig:threshold-distr}, \ref{fig:threshold-distr-tev}. Such an
energy dependent offset relative to the SM prediction should be
visible in Tevatron or LHC analyses.
To ellaborate on this point further it is again useful to study the
parametric dependence of the threshold cross section with respect to
$g_Y$ similar to what has been done in section \ref{sec-large-momenta} 
for the inclusive cross section. Restricting the cross section to the
threshold region by introducing a cut on the invariant mass of the
top-quark pair we find for the LHC:
\begin{eqnarray}
{\delta\sigma^{ 8\TeV}_{\rm EW} }(m_{t\bar t} < 2M_t +  50 \GeV) &=& ( -0.10  + 0.09 g_Y^2 ) \pb, \\
{\delta\sigma^{ 8\TeV}_{\rm EW} }(m_{t\bar t} < 2M_t + 100 \GeV) &=& ( -0.19  + 0.13 g_Y^2 ) \pb, \\
{\delta\sigma^{ 8\TeV}_{\rm EW} }(m_{t\bar t} < 2M_t + 150 \GeV) &=& ( -0.24  + 0.13 g_Y^2 ) \pb. 
\end{eqnarray}
Note that we neglected the tiny contribution linear in $g_Y$ since it
is irrelevant for the phenomenology.  For the relative corrections we
find:
\begin{eqnarray}
{\delta\sigma^{ 8\TeV}_{\rm EW} \over \sigma_{\LO}}(m_{t\bar t} < 2M_t +  50 \GeV) &=& ( -3.53  + 3.14 g_Y^2 )\%,  \\
{\delta\sigma^{ 8\TeV}_{\rm EW} \over \sigma_{\LO}}(m_{t\bar t} < 2M_t + 100 \GeV) &=& ( -3.05  + 2.05 g_Y^2 )\%,  \\
{\delta\sigma^{ 8\TeV}_{\rm EW} \over \sigma_{\LO}}(m_{t\bar t} < 2M_t + 150 \GeV) &=& ( -2.83  + 1.54 g_Y^2 )\%. 
\end{eqnarray}
Results for 13 and 14 TeV collider energy are given in section
\ref{sec:numresults}.  Using $m_{t\bar t} < 2M_t+50 \GeV$ we see, that
doubling the Yukawa coupling would lead to a change of the cross
section of about 9\%. As expected, restricting the cross section to
the threshold region significantly enhances the sensitivity to the
top-quark Yukawa coupling. Assuming an experimental precision of 5\%
the top-quark Yukawa coupling can be constrained.  The relevance of
such a limit on $g_Y$ can be illustrated by comparing with the limits
on the Higgs couplings as suggested in Refs.
\cite{Caola:2013yja,Kauer:2012hd} and presented recently by both CMS
\cite{Khachatryan:2014iha} and ATLAS \cite{atlas-higgs}
collaborations. In these papers it has been demonstrated that a
universal rescaling $g^{\rm SM}\to \kappa g^{\rm SM}$ combined with an
increase of the Higgs width by a factor $\kappa^4$ (e.g. through some
presently invisible mode) can be excluded for values of $\kappa^4$
exceeding 7.7 \cite{atlas-higgs} or even 5.4
\cite{Khachatryan:2014iha}. Such a rescaling would also involve the
top quark Yukawa coupling, and it remains to be seen, if similar
limits can be obtained from analysis of the top quark threshold
behaviour.

\section{Outlook and conclusions}
\label{outlook}
A sizable data sample has been collected by LHC experiments at a
center-of-mass energy of 8~TeV, and the Higgs boson has been discovered
with a mass of about 126~GeV. In view of these developments an update of
the weak corrections to top quark pair production has been
presented. We demonstrate that these corrections start to become
important already for the 8~TeV run, if an experimental precision of 5
percent can be reached. This observation applies both for large
transverse momenta, say above 500~GeV, where negative corrections around
5\% are observed, and for top quark production close to threshold which
is enhanced by about 5\% due to the attractive Yukawa interaction. A
detailed study of the top-antitop spectrum close to threshold could,
therefore, determine the strength of the Yukawa coupling or, at
least provide interesting upper limits. We also investigate the
distribution of top and antitop with respect to their rapidity
difference $\Delta y_{t\bar t}$ for the subsample with large invariant
mass and observe marked distortions of order 8\% (LHC8) and 12\%
(LHC14). Clearly these effects might be misinterpreted as evidence for 
anomalous couplings and thus have to be well under control. Last not
least we indicate a possible approach for combining QCD and weak
corrections in the framework of a Monte Carlo generator.

{\bf Acknowledgments}\\[3mm]
The work of J.H.K. was supported by BMBF Project 05H12VKE.
Discussions with Jeannine Wagner-Kuhr on experimental issues are
gratefully acknowledged. The work of P.U. was supported by BMBF
Project 05H12KHE.
The work of A.S. has been partly supported by
the German Ministry of Education and Research (BMBF) under contract
no. 05H12WWE.

\appendix
\section{Results for LHC operating at 13 or 14 TeV}
\subsection{Parametrization of 13 and 14 TeV cross section as function of
  the top-quark Yukawa coupling}
\label{sec:numresults}
\begin{eqnarray}
{\delta\sigma^{ 13\TeV}_{\rm EW} }(m_{t\bar t} < 2M_t +  50 \GeV) &=& ( -0.309  + 0.25 g_Y^2 ) \pb, \\
{\delta\sigma^{ 13\TeV}_{\rm EW} }(m_{t\bar t} < 2M_t + 100 \GeV) &=& ( -0.572  + 0.36 g_Y^2 ) \pb, \\
{\delta\sigma^{ 13\TeV}_{\rm EW} }(m_{t\bar t} < 2M_t + 150 \GeV) &=& ( -0.752  + 0.38 g_Y^2 ) \pb,
\end{eqnarray}
\begin{eqnarray}
{\delta\sigma^{ 14\TeV}_{\rm EW} }(m_{t\bar t} < 2M_t +  50 \GeV) &=& ( -0.360  + 0.29 g_Y^2 ) \pb, \\  
{\delta\sigma^{ 14\TeV}_{\rm EW} }(m_{t\bar t} < 2M_t + 100 \GeV) &=& ( -0.669  + 0.41 g_Y^2 ) \pb,  \\ 
{\delta\sigma^{ 14\TeV}_{\rm EW} }(m_{t\bar t} < 2M_t + 150 \GeV) &=& ( -0.879  + 0.44 g_Y^2 ) \pb. 
\end{eqnarray}
The relative corrections are given by:
\begin{eqnarray}
  {\delta\sigma^{ 13\TeV}_{\rm EW} \over \sigma_{\LO}}(m_{t\bar t} < 2M_t +  50 \GeV) &=& ( -3.753  + 3.08 g_Y^2 )\%,  \\
  {\delta\sigma^{ 13\TeV}_{\rm EW} \over \sigma_{\LO}}(m_{t\bar t} < 2M_t + 100 \GeV) &=& ( -3.177  + 1.97 g_Y^2 )\%,  \\
  {\delta\sigma^{ 13\TeV}_{\rm EW} \over \sigma_{\LO}}(m_{t\bar t} <
  2M_t + 150 \GeV) &=& ( -2.908  + 1.45 g_Y^2 )\%,  
\end{eqnarray}
\begin{eqnarray}
  {\delta\sigma^{ 14\TeV}_{\rm EW} \over \sigma_{\LO}}(m_{t\bar t} <
  2M_t +  50 \GeV) &=& ( -3.775  + 3.08 g_Y^2 ) \%, \\  
  {\delta\sigma^{ 14\TeV}_{\rm EW} \over \sigma_{\LO}}(m_{t\bar t} < 2M_t + 100 \GeV) &=& ( -3.190  + 1.96 g_Y^2 )\%,   \\
  {\delta\sigma^{ 14\TeV}_{\rm EW} \over \sigma_{\LO}}(m_{t\bar t} < 2M_t + 150 \GeV) &=& ( -2.914  + 1.44 g_Y^2 )\%.  
\end{eqnarray}
\subsection{Differential distributions for LHC operating at 13 and 14 TeV}
\label{sec:13and14TeVresults}
\begin{figure}
  \begin{center}
    \includegraphics[width=0.6\textwidth]{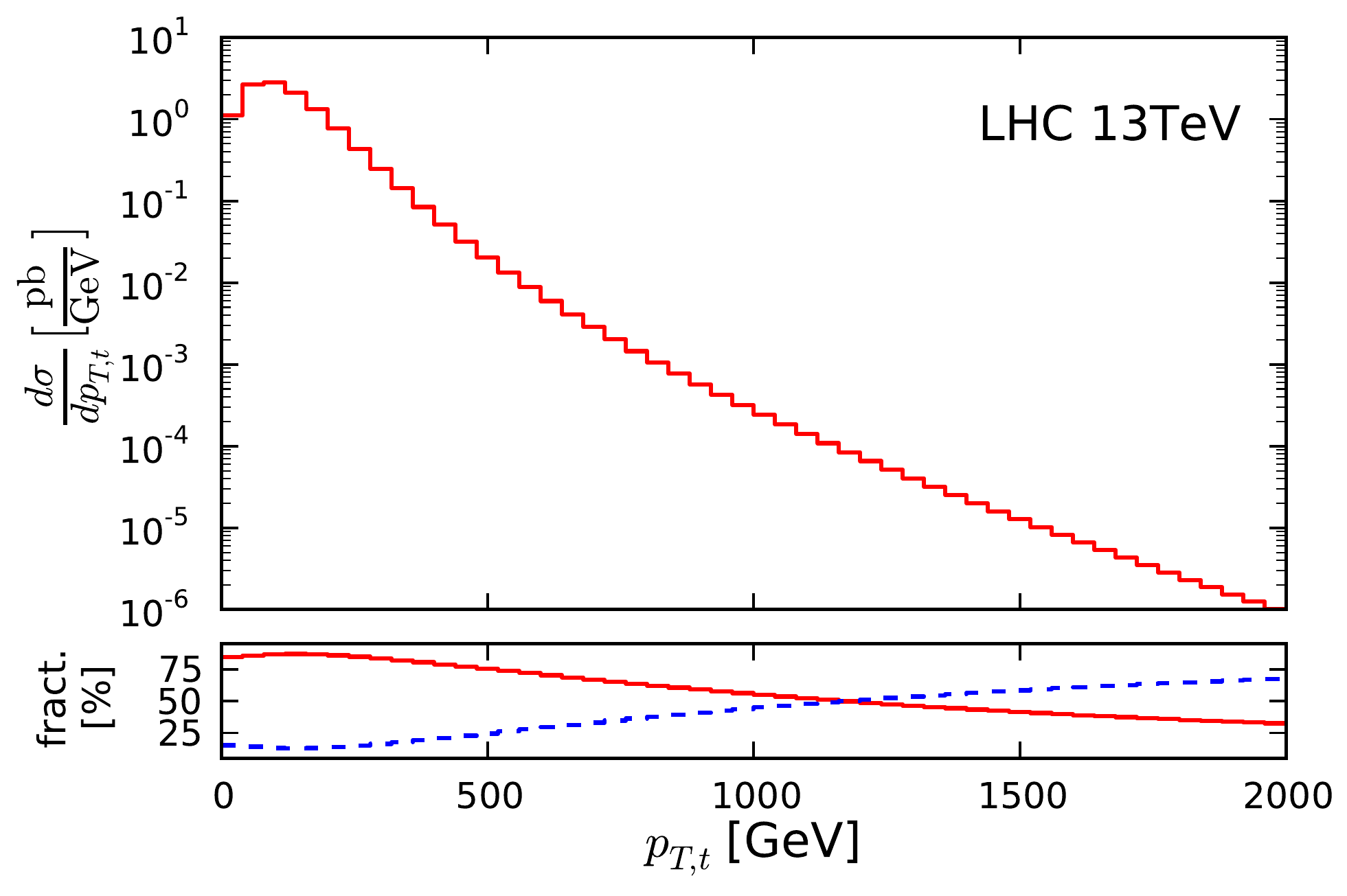}
    \caption{Same as Fig.~\ref{fig:LHC8-pt} but for 13 TeV.}
  \end{center}
\end{figure}
\begin{figure}
  \begin{center}
    \includegraphics[width=0.6\textwidth]{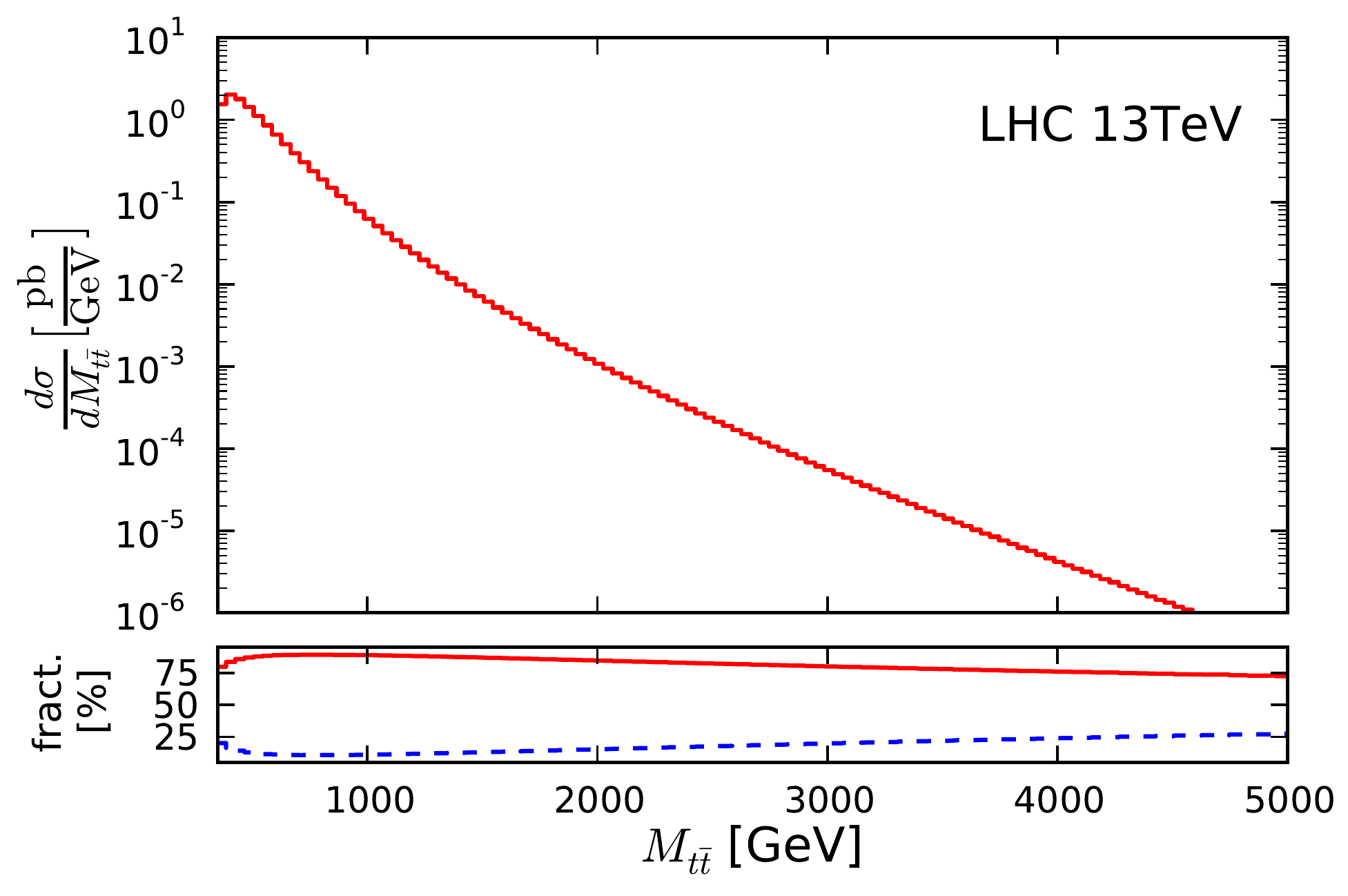}
    \caption{Same as Fig.~\ref{fig:LHC8-mtt} but for 13 TeV.
      }
  \end{center}
\end{figure}\begin{figure}
  \begin{center}
    \includegraphics[width=0.6\textwidth]{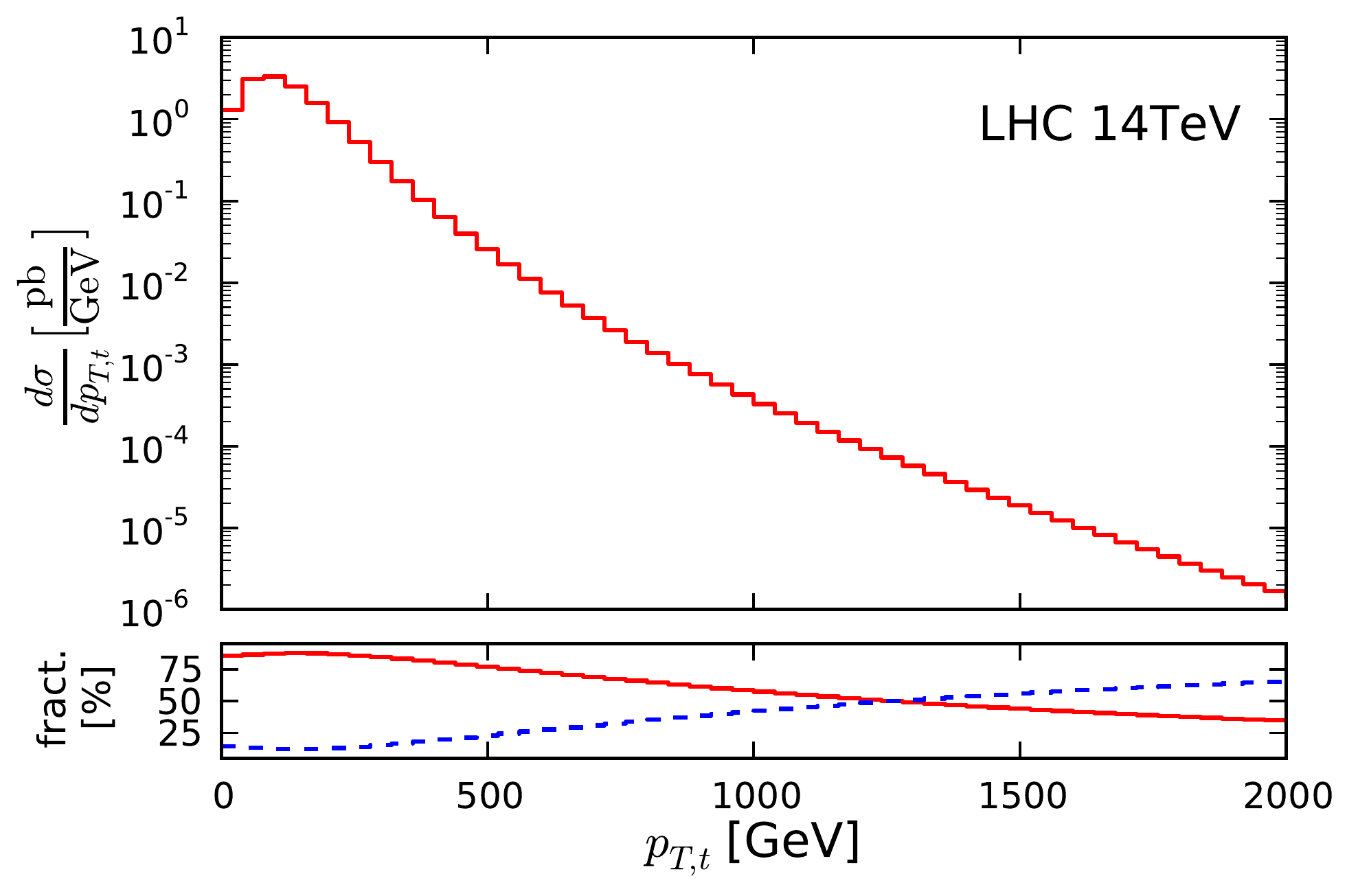}
    \caption{Same as Fig.~\ref{fig:LHC8-pt} but for 14 TeV.}
  \end{center}
\end{figure}
\begin{figure}
  \begin{center}
    \includegraphics[width=0.6\textwidth]{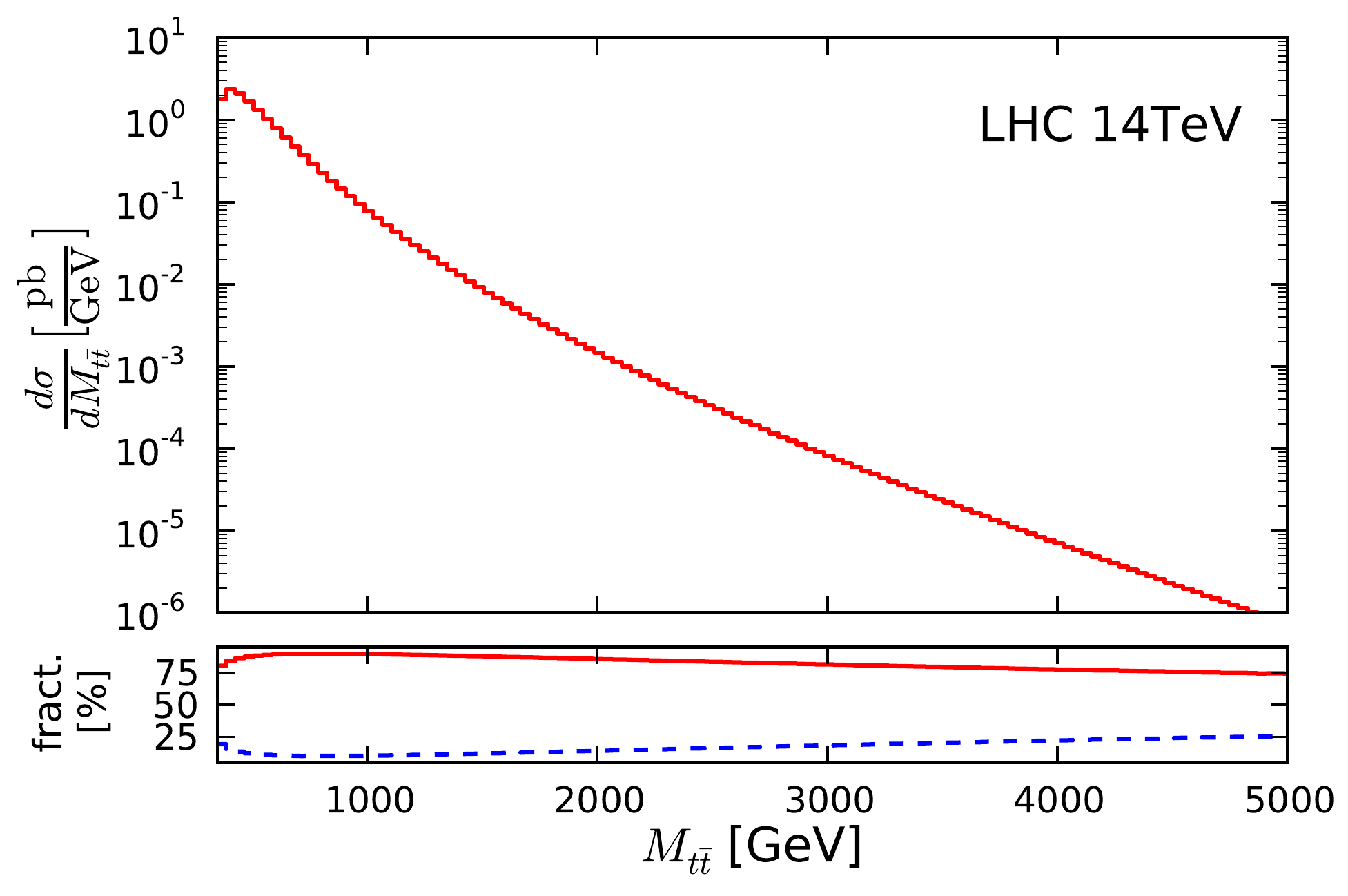}
    \caption{Same as Fig.~\ref{fig:LHC8-mtt} but for 14 TeV.
      }
  \end{center}
\end{figure}

\begin{figure}
  \begin{center}
    \includegraphics[width=0.5\textwidth]{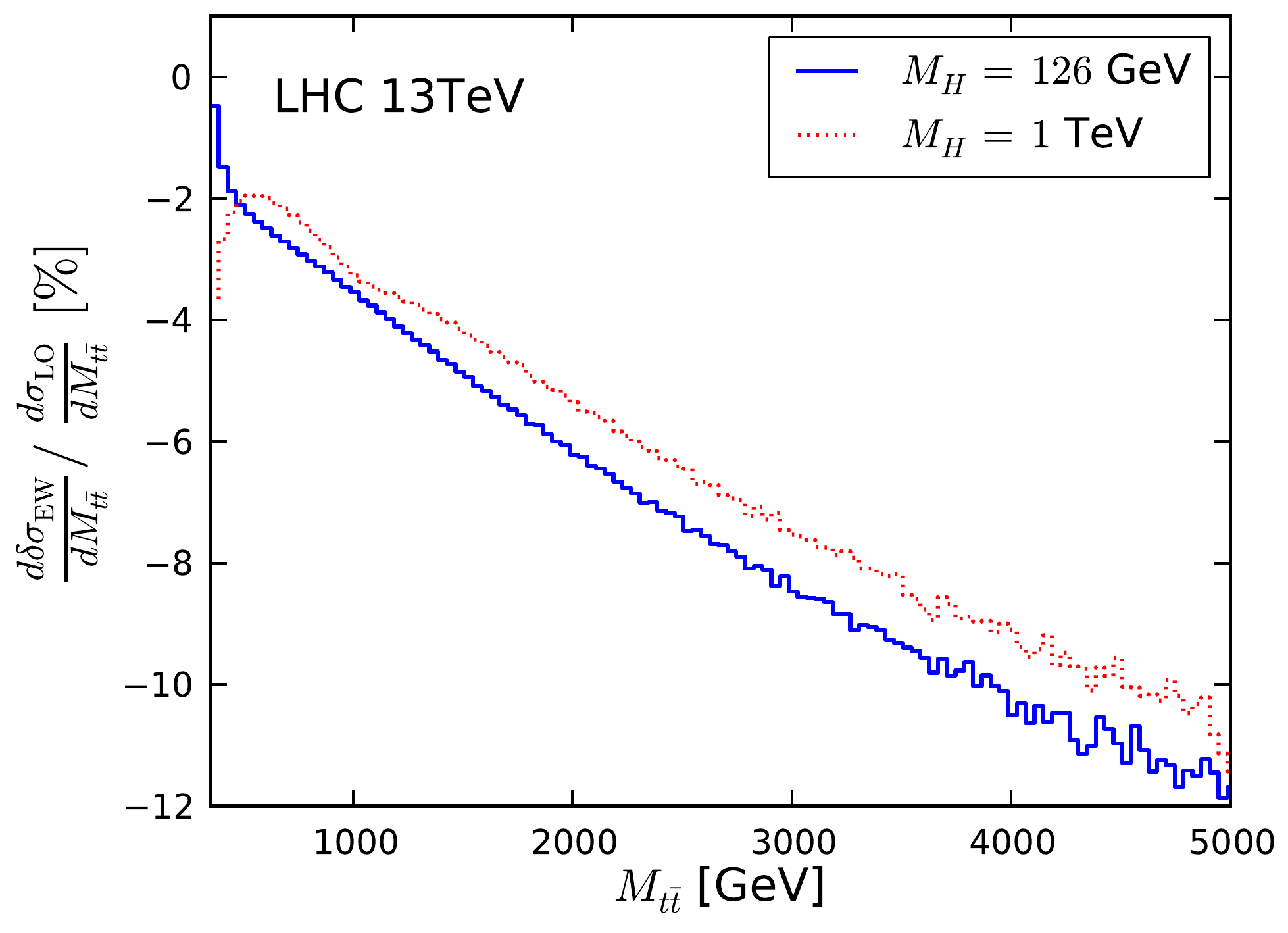}
    \includegraphics[width=0.48\textwidth]{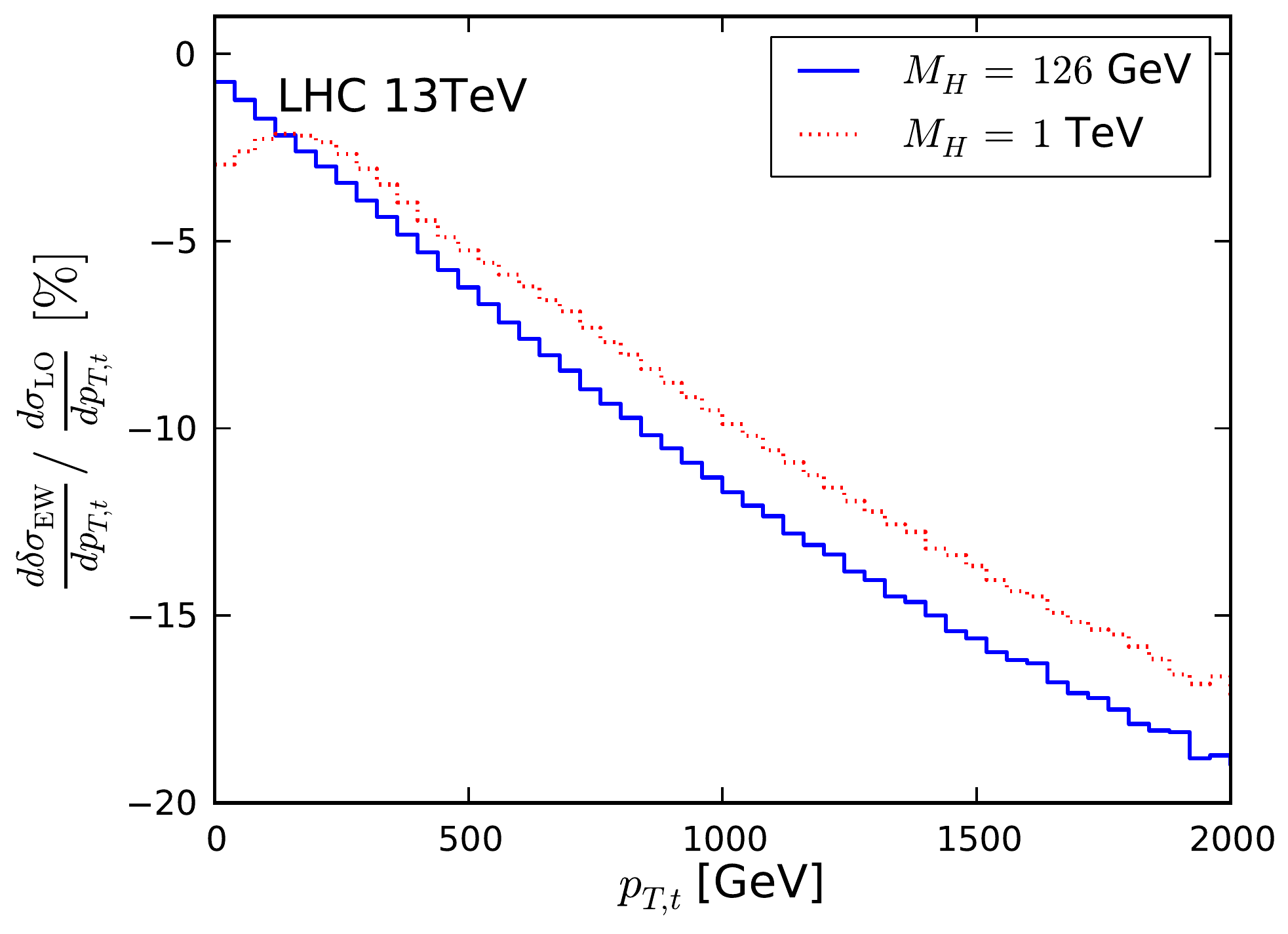}
 \caption{Same as Fig.~\ref{fig:distr8} but for 13 TeV.
}
     \end{center}
\end{figure}
\begin{figure}
  \begin{center}
    \includegraphics[width=0.5\textwidth]{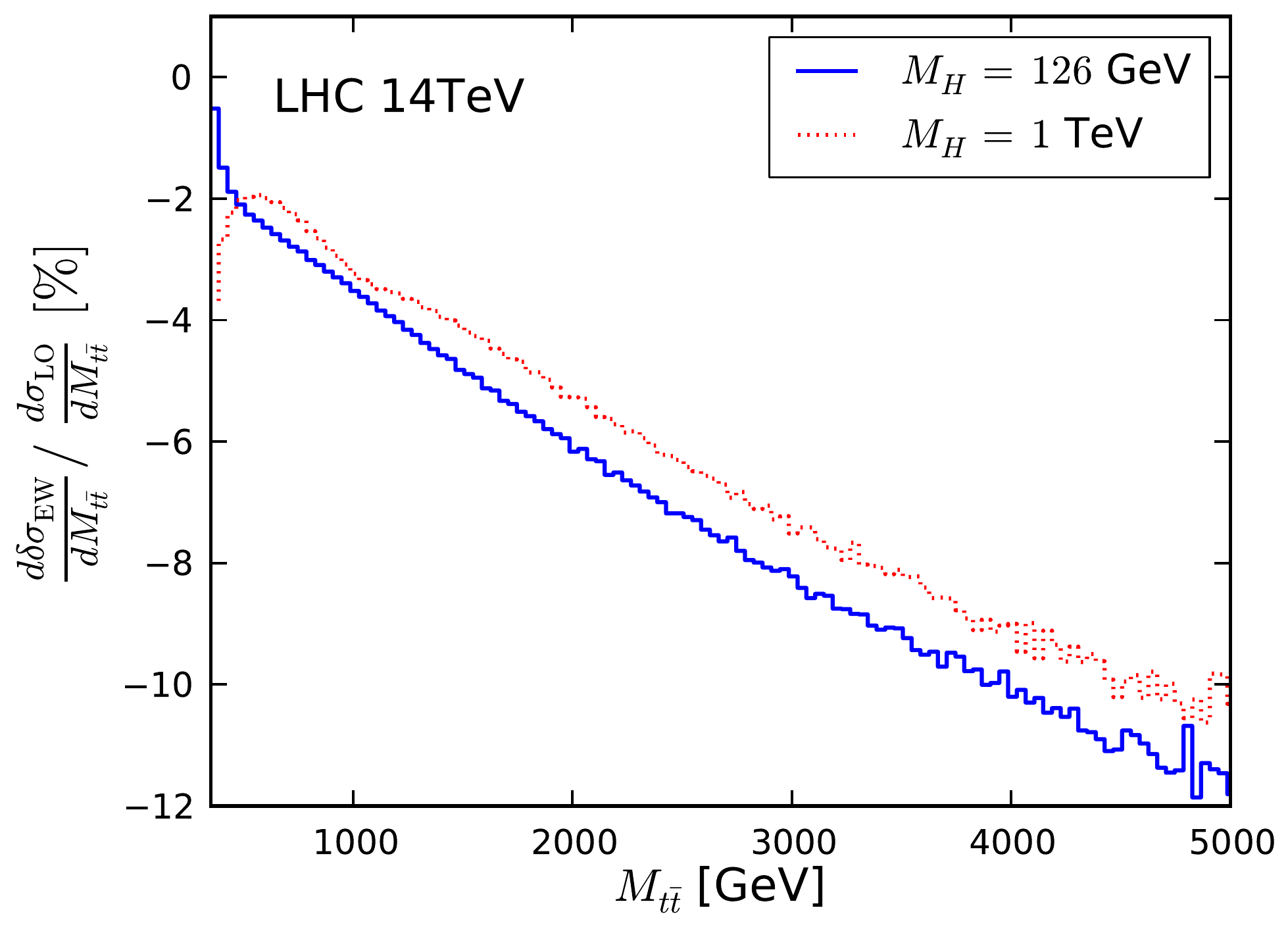}
    \includegraphics[width=0.48\textwidth]{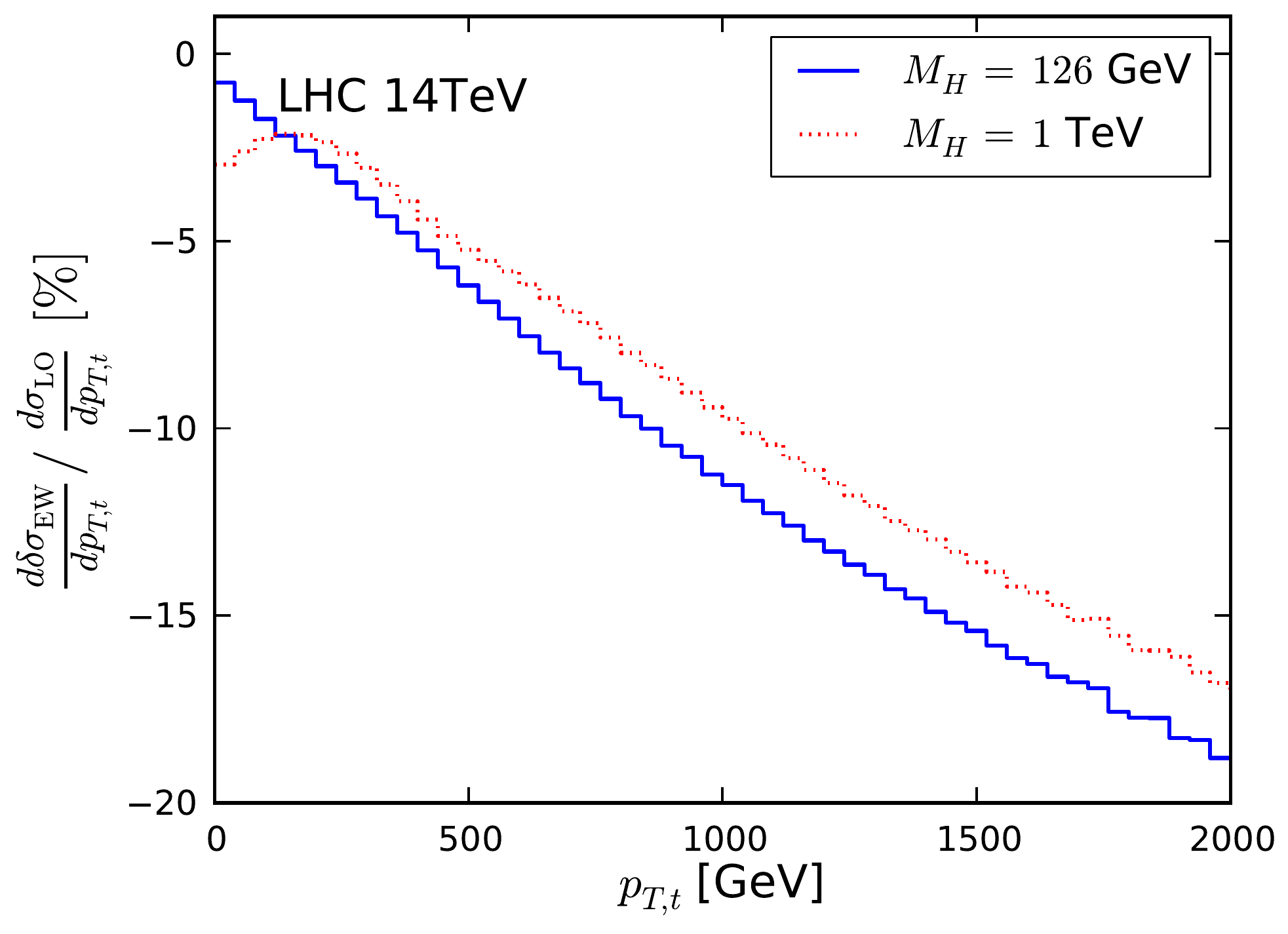}
 \caption{Same as Fig.~\ref{fig:distr8} but for 14 TeV.
}
     \end{center}
\end{figure}
\begin{figure}
  \begin{center}
    \includegraphics[width=0.45\textwidth]{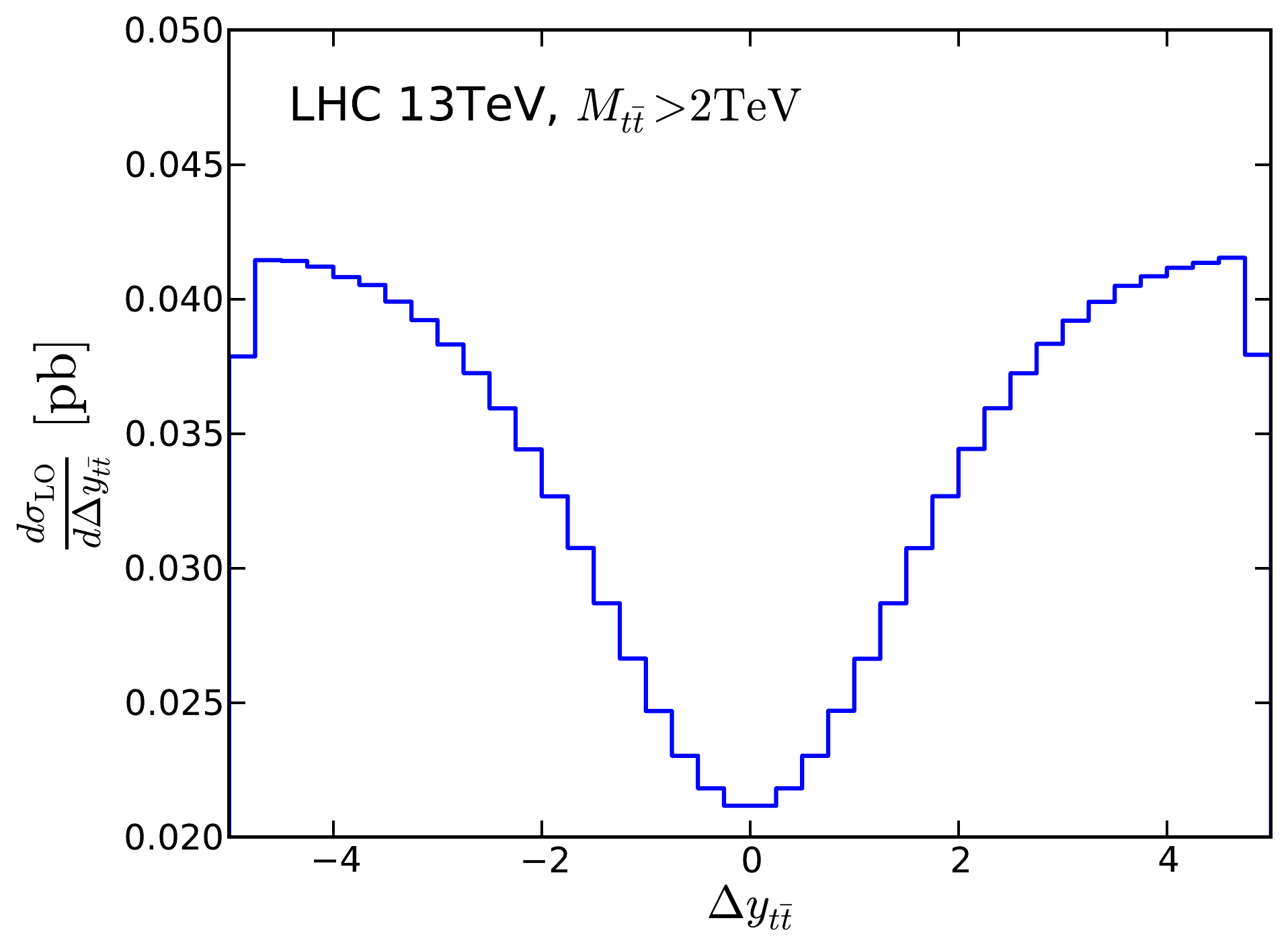}
    \includegraphics[width=0.45\textwidth]{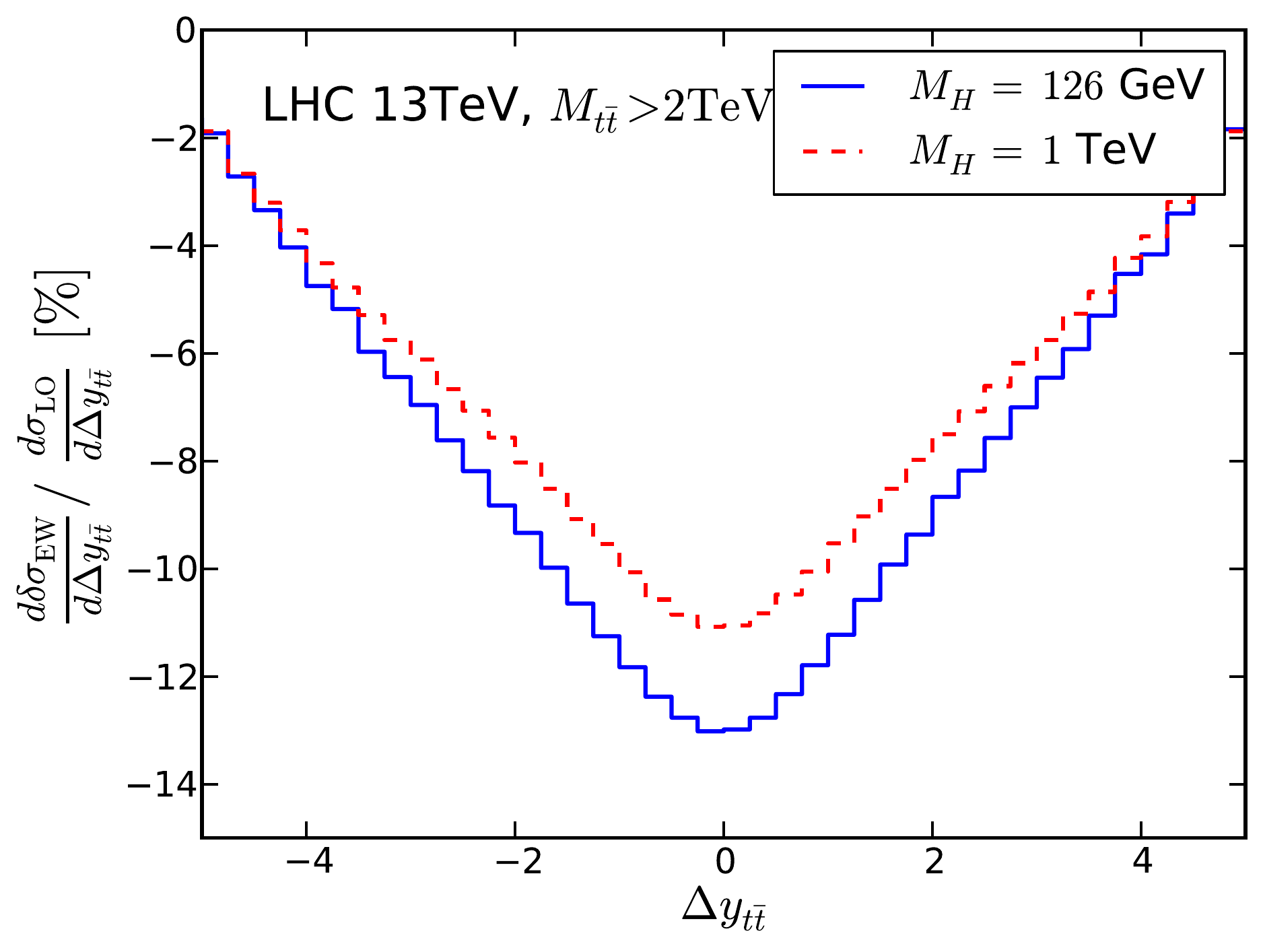}
 \caption{Same as Fig.~\ref{fig:rapidity-distr} 
   but for 13 TeV.}
     \end{center}
\end{figure}   
\begin{figure}
  \begin{center}
    \includegraphics[width=0.45\textwidth]{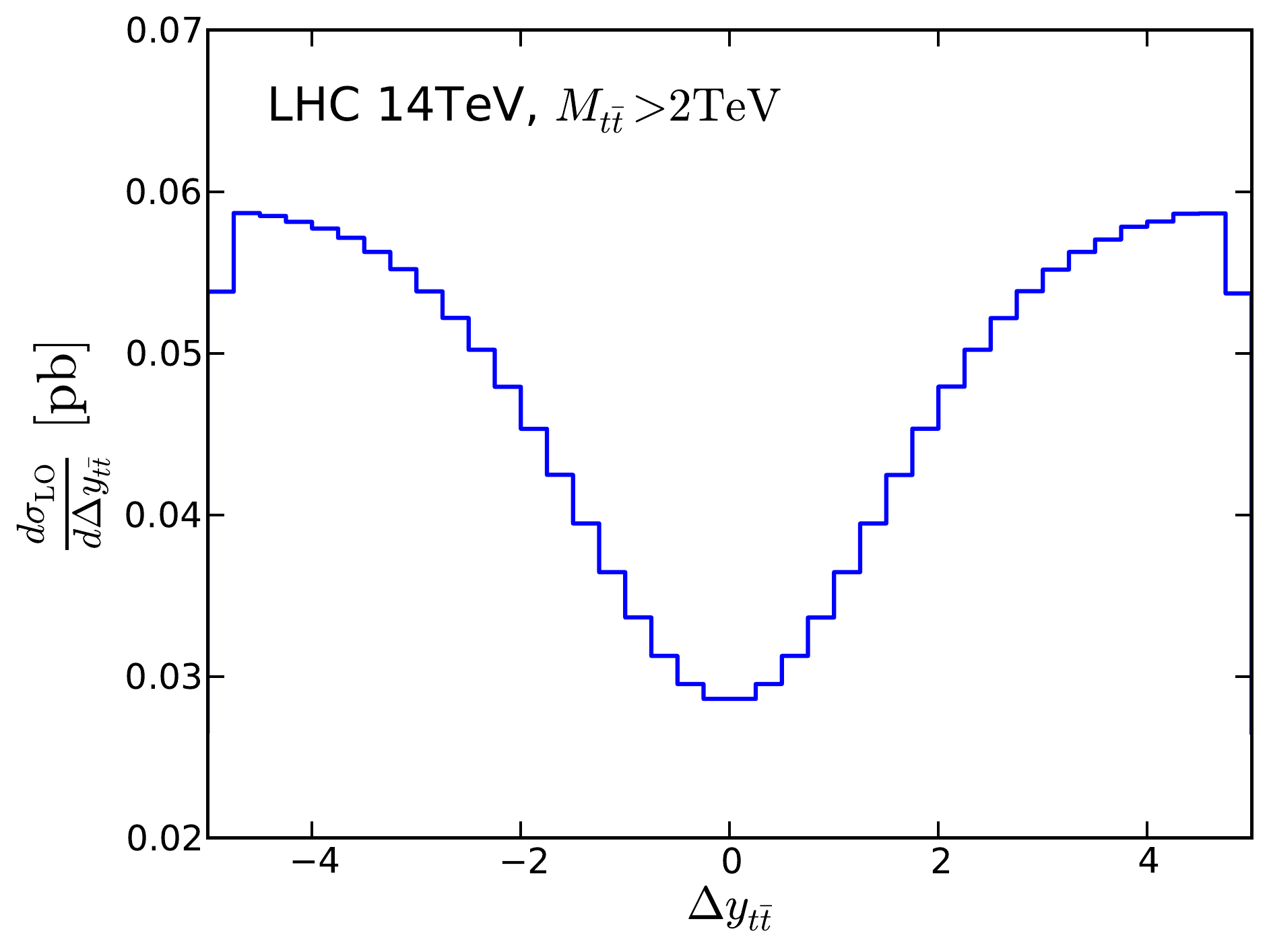}
    \includegraphics[width=0.45\textwidth]{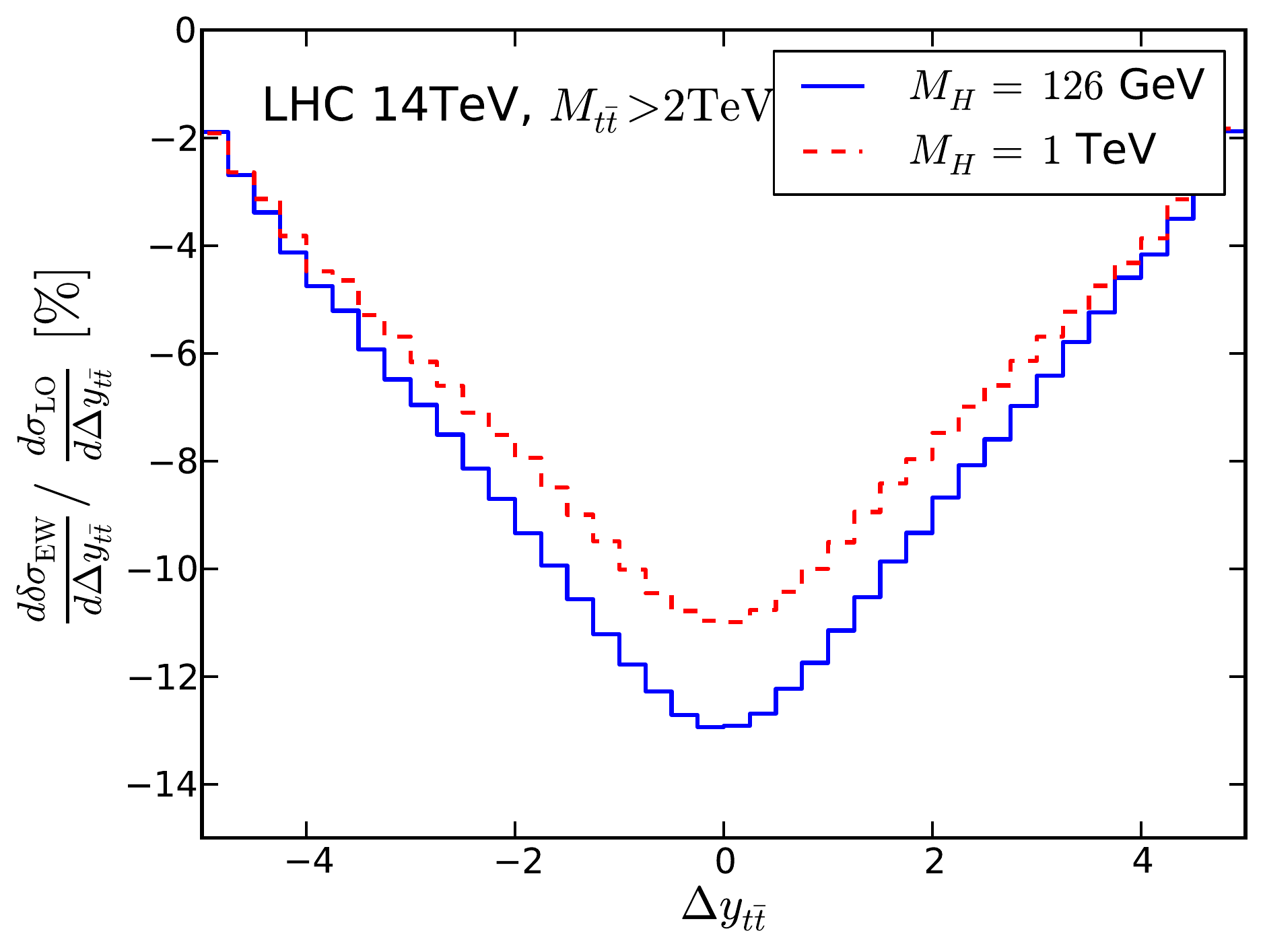}
 \caption{Same as Fig.~\ref{fig:rapidity-distr} 
   but for 13 TeV.}
     \end{center}
\end{figure}   
\begin{figure}
  \begin{center}
    \includegraphics[width=0.6\textwidth]{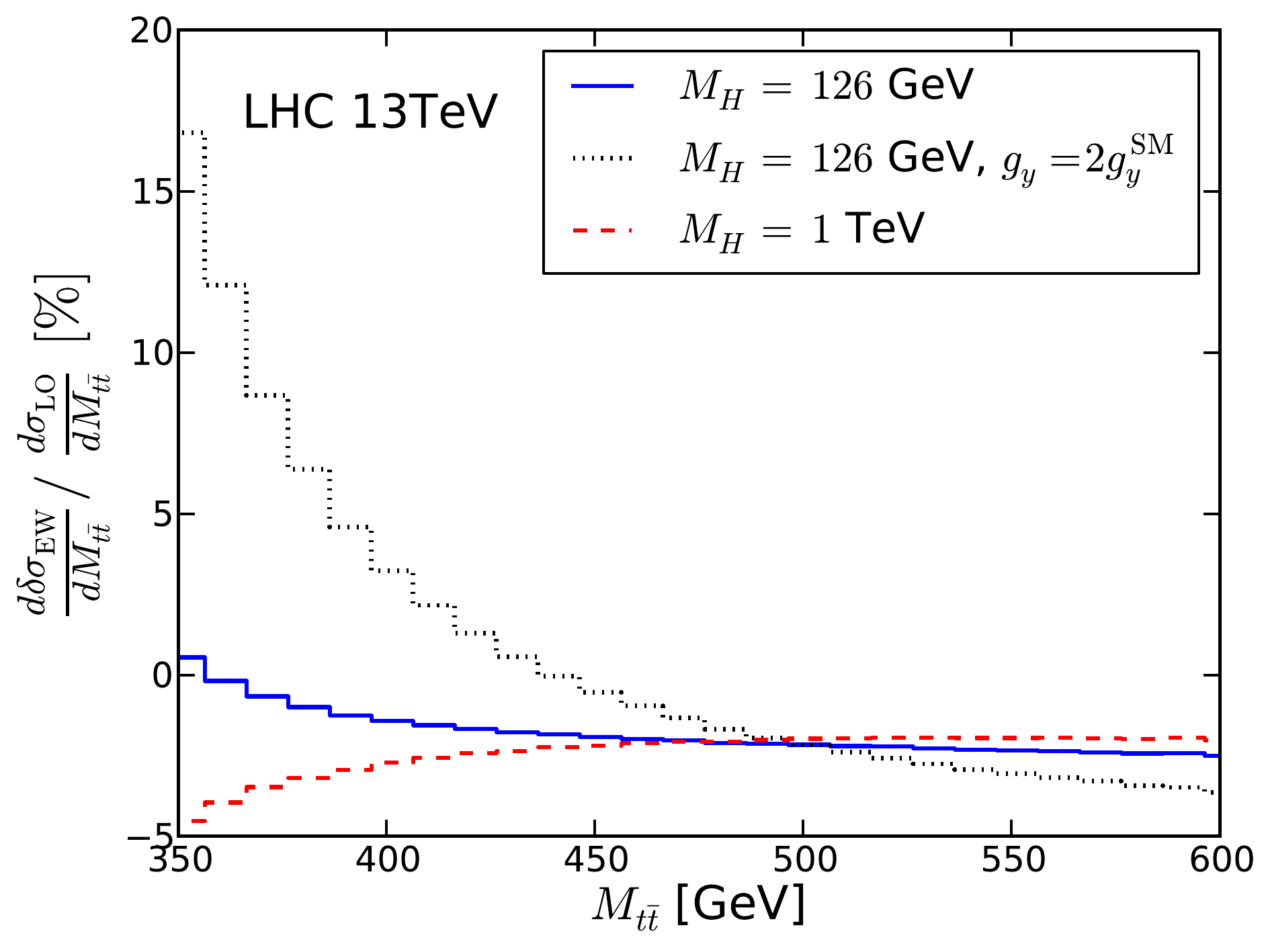}
    \caption{Same as Fig.~\ref{fig:threshold-distr} but for 13 TeV.
      }
  \end{center}
\end{figure}   
\begin{figure}
  \begin{center}
    \includegraphics[width=0.6\textwidth]{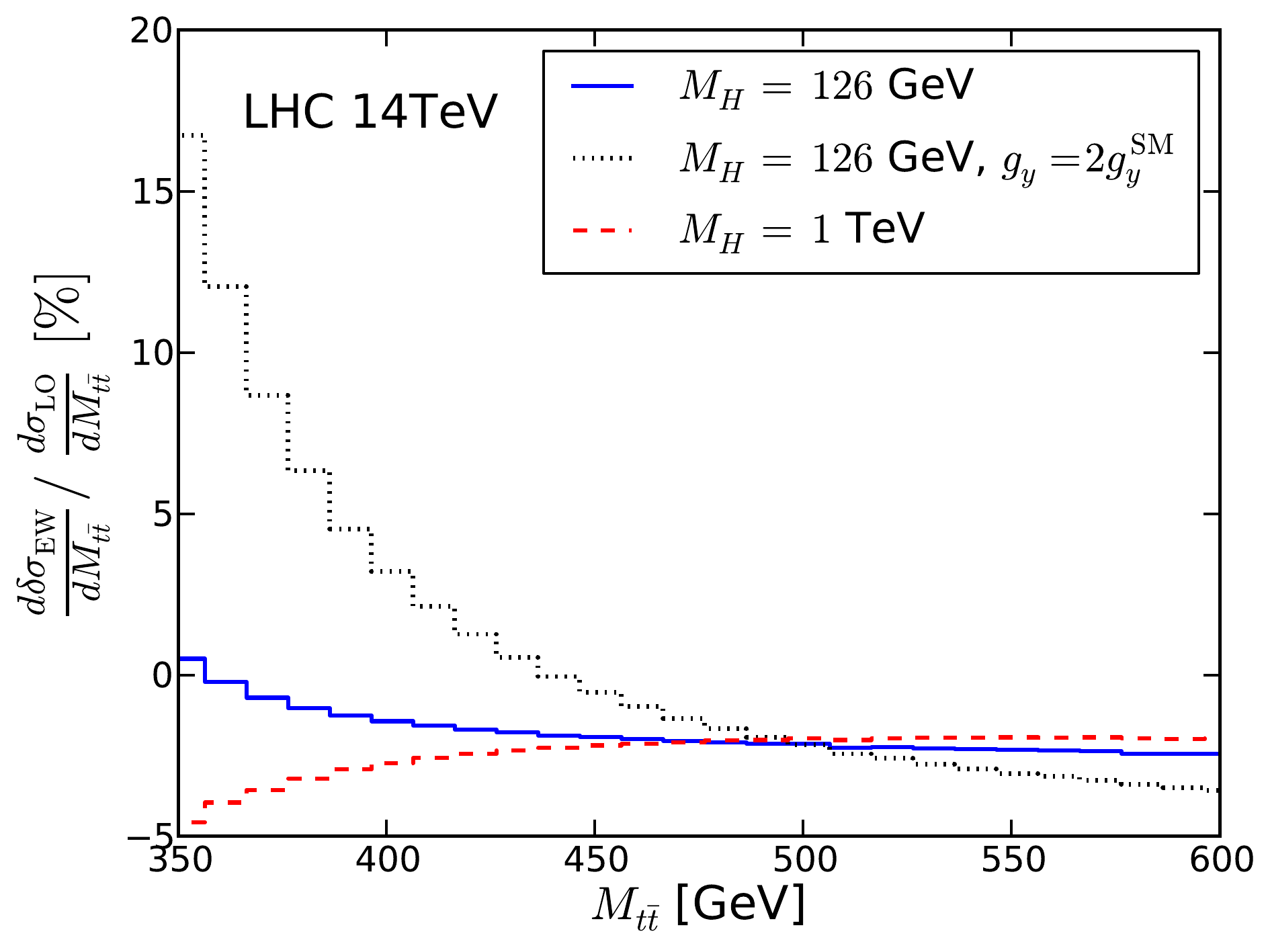}
    \caption{Same as Fig.~\ref{fig:threshold-distr} but for 14 TeV.
      }
  \end{center}
\end{figure}

\end{document}